\documentclass[11.0pt,a4paper]{article}
\usepackage{epsfig}

\newlength{\dinwidth}
\newlength{\dinmargin}
\def\ytau{y}

\def\eq#1{{Eq.~(\ref{#1})}}
\newcommand{\Le}{\left(}
\newcommand{\Ra}{\right)}

\newcommand{\al}{\alpha}
\newcommand{\T}{T}
\newcommand{\beq}{\begin{equation}}
\newcommand{\eeq}{\end{equation}}
\newcommand{\beqar}[1]{\begin{eqnarray}\label{#1}}
\newcommand{\eeqar}{\end{eqnarray}}
\begin{document}

\title {{~}\\
{\Large \bf  Gluon density and $F_{2}$ functions from BK equation
with local impact parameter dependence in DIS on nuclei}\\}
\author{ 
{~}\\
{~}\\
{\large 
S.~Bondarenko$\,{}^{a,b)}\,$\thanks{Email: sergeyb@ariel.ac.il}}
 \\[10mm]
{\it\normalsize $^{a)}$ Leaving University Santiago De Compostela, Spain.}\\
{\it\normalsize $^{b)}$ Ariel University, Israel.}\\}

\maketitle
\thispagestyle{empty}

\begin{abstract}
The DIS process on nuclei is considered in the framework of 
LO BK equation with local impact parameter dependence. 
The initial conditions of GBW type,
found in \cite{Serg1}, was used in the solution of BK equation. Integrated
gluon density function and  $F_2$ nucleus structure function
for different nuclei are calculated. Obtained results
are compared  with the different parameterizations of integrated gluon density function 
from \cite{Flor,Eskola1,Eskola2,Tywon}. 
The anomalous dimensions and saturation scales for different nuclei 
are calculated at different energies.  The expressions of the
functional form of saturation scale are obtained for the proton and different nuclei as functions 
of the impact parameter and energy.
\end{abstract}

\section{Introduction}

\,\,The phenomenological applications of the QCD BFKL Pomeron, \cite{bfkl},
and of the system of interacting Pomerons, \cite{braun1,braun2},
could be investigated in the framework of the BK equation
\cite{bal,kov}. This framework was used in the
calculations of the amplitudes of the different scattering processes, 
see \cite{bom4,bom,pryl, bom1,Bond,bom2} for example and references therein.
The same principals of the saturation physics were applied 
in the different phenomenological and CGC type models as well,
see \cite{GolecBiernat1,GolecBiernat2,GolecBiernat3,Mueller1,
Munier1,Lev1,Lev2,Lev3,Bartels1,Kowalski1,Iancu1,Alb,Kowalski2}.
In these approaches, inspired by BK equation,
the impact parameter dependence of the 
solution of the equation was
treated only approximately, see for example \cite{Lev11,Lev2}, whereas the 
phenomenological
models, such as  GBW model \cite{GolecBiernat1,GolecBiernat2},
treating the impact parameter dependence
neglect the evolution of the amplitude with rapidity.
There are also the studies of the
impact parameter dependence of gluon structure function together with
DGLAP evolution of the function considered in the papers 
\cite{Kwi1,GolecBiernat3,Kowalski1,Kowalski2} and calculations of the unintegrated 
gluon density function in the framework of modified BK equation
with the factorized dependence of impact parameter and momentum in
initial conditions in \cite{Kut1,Kut2,Kut3}.
Nevertheless, the investigation of the BK equation with the new type
of the initial conditions
with the non-factorized impact parameter and momentum dependence 
is a task which requires  a deep attention due the importance of the
BK equation in the different scattering processes with nuclei involved.

 In the paper \cite{Serg1} a first step towards the including of the impact 
parameter dependence into the rapidity evolution of the scattering amplitude
through initial conditions was made. It must be stressed, 
that introducing local impact parameter dependence into evolution equation
trough initial condition
we miss the precise treatment of transverse position in the evolution
kernel, that does not allow to calculate the contribution of the Pomeron loops into the 
amplitude, for example. Nevertheless, considered  semi-classical framework with local 
impact parameter dependence is justified due the following observation. 
For the scattering on the target, which size is large, the momentum transfer
of the Pomeron line originating from the external particle is bounded
by the form-factors of the sources, and, therefore, it is small for the nuclei (and possibly proton) targets.
Therefore, when we consider only "net" diagram structure, 
i.e. the semi-classical approximation to the problem, and when we do not 
account the Pomeron loops contribution, the constraint on the momentum transfer
of the Pomeron line  is imposed on the all Pomeron lines that justify
our approximation, see more details in \cite{bom} and references therein.

 The DIS process on the proton, which was considered in the paper \cite{Serg1},
allowed to find a initial condition for the BK evolution by the fitting 
of the $F_2$ function data. 
In the present studies we use the same as in \cite{Serg1} calculation procedure
solving  LO BK equation
in each point of the impact parameter space, using methods
developed in \cite{bom}. The difference with the calculations from \cite{Serg1}
is that now we consider different nuclei as a target instead the 
proton target in the paper \cite{Serg1}, and, therefore, 
the impact parameter profile of the proton we change on the Wood-Saxon 
parameterization of the nuclei density profile. 
All other parameters
of the initial profile for BK equation we take the same, 
in spite of the differences of the processes. 
The applicability of this assumption we will discuss 
latter. As it will be shown,  the change of only impact parameter 
profile in initial conditions will lead to the results for integrated gluon density 
function similar to the results of the different calculations of the same function in 
\cite{Flor,Eskola1,Eskola2,Tywon}, justifying this
minimal modifications of the  initial conditions for DIS on nuclei.

 The comparison of our results with the results of other calculations is
based on the fact, that 
due the lack of the high energy DIS process data for the nuclei targets, 
we could not perform the fitting of the data in order to determine all
parameters for initial conditions as it was done in  \cite{Serg1}.
More of that, we also could not use the low energy data for this purpose, 
because the small Bjorken x evolution begins at the energies larger than 
the energies for
which the nuclei data are known. Fortunately, the existing
models of the integrated gluon density,  \cite{Flor,Eskola1,Eskola2,Tywon}, 
allow to extrapolate the integrated
gluon density function till very small values of x, $x\approx\,10^{-7}$.
It gives to us a possibility to check how good (or bad) our 
calculations are and, therefore, how good (or bad) our initial conditions are.
To our surprise, curves for the integrated gluon density of different nuclei
obtaining in our framework are in the "window" of
the curves from the \cite{Flor,Eskola1,Eskola2,Tywon}
which were calculated in different approaches. This result, as we underlined previously, 
justify the use of "minimal changed" initial conditions
for the nuclei in our calculations.

 The knowledge of the unintegrated gluon density function, i.e. solution of 
BK equation,
allows to find other functions and parameters connected with the DIS process.
We calculate the integrated gluon density function and 
$F_2$ structure function of the DIS process on nuclei, or, more precisely
the proton-nuclei ratio of these functions. In this case we find a 
parameterization of the nuclei integrated gluon density and $F_2$ functions in the terms of the corresponding
proton functions and number of nucleons in the nuclei.
We also calculate a anomalous dimension of the 
integrated gluon density function, similarly to the calculations of \cite{Ayala1},
and find a saturation momenta of the process as a function of impact parameter space
and energy.

 This paper is organized as follows. In the next section we describe a 
formalism of the calculations. In Sec.3 we present obtained results for the 
integrated gluon density function.  In Sec.4 we consider the calculations of the 
$F_2$ structure function and in Sec.5 we present results for the anomalous dimensions 
of the integrated gluon density function of the DIS process on the proton and on the nuclei.
The saturation scale calculations for the problem of DIS process on the proton
and DIS process on the nuclei are presented in the Section 6.
Section 7 is a conclusion of the paper.

\section{The low-x structure function in the momentum representation }

 In this section we shortly remind the main formulae used in the following calculations
(see also \cite{Serg1}).
The unintegrated gluon density function $f(x,\,k^2,b)$ of the DIS process 
we find solving  BK
equation for the each point in the impact parameter space:
\[
\partial_\ytau f(\ytau,k^2,b) =
{N_c \alpha_s \over \pi}\, k^2\, \int {da^2 \over a^2}
\left[
{f(a^2,b)-f(k^2,b) \over |a^2-k^2|} +
{f(k^2,b)\over [4a^4+k^4]^{{1\over 2}}}
\right]
\]
\beq\label{F2}
- {2\pi \alpha_s ^2} \;
\left[
k^2 \int_{k^2} {da^2 \over a^4} \;  f(a^2,b)
\int_{k^2} {dc^2 \over c^4} \;  f(c^2,b)
+ f(k^2,b)\int_{k^2} {da^2 \over a^4} \;
\log\left( {a^2 \over k^2} \right) f(a^2,b)\right]
\eeq
where we introduced the rapidity variable $y=\log(1/x)$.
In Eq.\ref{F2} we assumed, that the evolution is local in the
transverse plane, i.e. impact parameter dependence
of $f(\ytau,k^2,b)$ appear only throw the initial condition
for the $f(\ytau,k^2,b)$ function
\beq\label{F3}
f(\ytau\,=\,\ytau_{0}\,,k^2,b)\,=\,f_{in}(k^2,b)\,
\eeq
In order to exclude part of ambiguities in the solution of BK equation which arise
due the absence of the  NLO corrections, we perform the following
substitute in the equation
\beq\label{F4}
f(\ytau,k^2,b)\rightarrow\,\frac{f(\ytau,k^2,b)\alpha_s(k^2)}{\alpha_s}=
\frac{\tilde{f}(\ytau,k^2,b)}{\alpha_s}\,,
\eeq
obtaining
\[
\partial_{\tilde{\ytau}} \tilde{f}(\tilde{\ytau},k^2,b) =
{N_c \over \pi}\, k^2\, \int {da^2 \over a^2}
\left[
{\tilde{f}(a^2,b)-\tilde{f}(k^2,b) \over |a^2-k^2|} +
{\tilde{f}(k^2,b)\over [4a^4+k^4]^{{1\over 2}}}
\right]
\]
\beq\label{F5}
- {2\pi} \;
\left[
k^2 \int_{k^2} {da^2 \over a^4} \;  \tilde{f}(a^2,b)
\int_{k^2} {dc^2 \over c^4} \;  \tilde{f}(c^2,b)
+ \tilde{f}(k^2,b)\int_{k^2} {da^2 \over a^4} \;
\log\left( {a^2 \over k^2} \right) \tilde{f}(a^2,b)\right]
\eeq
where $\tilde{y}=\alpha_s\,y$. The  value of $\alpha_s$ is a constant
in the LO approximation and
we consider $\alpha_s$ as parameter of the model, which
we borrow from the fit of DIS data performed in \cite{Serg1}. 

 The impact factors, used latter in calculations of $F_2$ structure function,
are usual impact factors of the problem with the
three light quarks flavors of equal mass included.
They have the following form (see \cite{Kwi1} for example):
\beq\label{F6}
\Phi_{L}(k,m_{q}^{2})\,=\,32\,\pi\,\al\,
\sum^{3}_{q=1}\,e^{2}_{q}\,
\int_{0}^{1}\,d\rho d\eta \,\,
\frac{k^2\,\eta\,(1-\eta)\,\rho^{2}\,(1-\rho)^{2}\,Q^{2}\,}
{\Le\,Q^{2}\rho(1-\rho)+k^{2}\eta(1-\eta)+m_{q}^{2}\Ra\,
\Le\,Q^{2}\rho(1-\rho)+m_{q}^{2}\Ra\,}
\eeq
and
\[
\Phi_{\T}(k,m_{q}^{2})\,=\,4\,\pi\,\al\,
\sum^{3}_{q=1}\,e^{2}_{q}\,
\int_{0}^{1}\,d\rho d\eta \,\cdot\,
\]
\beq\label{F7}
\,\cdot\,\frac{k^2\,Q^{2}\Le\,\rho^2 +(1-\rho)^{2}\,\Ra\,\rho\,
(1-\rho)\Le\eta^2 +(1-\eta)^{2}\,\Ra\,+
k^2\,\Le \rho^2 +(1-\rho)^{2}\,\Ra\,m_{q}^{2}\,+
4\,\rho\,(1-\rho)\,\eta(1-\eta)\,m_{q}^{2}}{
\Le\,Q^{2}\rho(1-\rho)+k^{2}\eta(1-\eta)+m_{q}^{2} \Ra\,
\Le\,Q^{2}\rho(1-\rho)+m_{q}^{2}\Ra\,}
\eeq
Due the including light quark masses in the calculations, the
rapidity variable $y$ (Bjorken x) in BK equation is also modified,
see details in \cite{GolecBiernat1,GolecBiernat2}.
For each fixed rapidity $y$ of the process, the value rapidity taken in BK equation
is changed
\beq\label{F9}
y\,\rightarrow\,y\,-\,\ln (1\,+\,\frac{4\,m^{2}_{q}}{Q^2})
\eeq

The form of the function $\tilde{f}(\ytau,k^2,b)$ at initial rapidity,
i.e. initial condition for the BK equation Eq.\ref{F5},
has been borrowed from the form of GBW ansatz,
\cite{GolecBiernat1,GolecBiernat2}, with introduced
impact parameter dependence. In the paper \cite{Serg1} 
the following form of the initial conditions for the DIS process on the proton 
was used 
\beq\label{F10}
\tilde{f}_{proton}(\ytau\,=\,\ytau_{0},k^2,b)\,=\,\frac{3}{4\pi^2\,}\,
\frac{k^4}{Q_{S}^{2}(b)}
exp(-\,\frac{k^2}{Q_{S}^{2}}\,)\,,
\eeq
where a saturation scale as a function of the impact parameter is defined 
as following
\beq\label{F11}
Q_{S}^{2}(b)\,=\,\frac{S(b)}{C}=\frac{e^{-b^2/R_{p}^{2}}}
{C\,\pi\,R_{p}^{2}}\,.
\eeq
The $S(b)$ function here is the proton impact parameter profile
with the usual normalization properties
\beq\label{F12}
\int\,S(b)\,d^{2}b\,=\,1\,
\eeq
and $C$ is a numerical coefficient which defines a value of the
saturation scale at zero impact parameter and initial rapidity 
through the proton radius
\beq\label{F13}
Q_{S}^{2}(b=0)\,=\,\frac{1}
{C\,\pi\,R_{p}^{2}}\,.
\eeq
The generalization of this initial condition for the case of nuclei is
straightforward
\beq\label{F14}
\tilde{f}_{nucleus}(\ytau\,=\,\ytau_{0},k^2,b)\,=\,\frac{3}{4\pi^2\,}\,
\frac{k^4}{Q_{SA}^{2}(b)}
exp(-\,\frac{k^2}{Q_{SA}^{2}}\,)\,,
\eeq
where 
\beq\label{F15}
Q_{SA}^{2}(b)\,=\,\frac{A\,S(b)}{C}\,
\eeq
with the  Wood-Saxon nuclei profile function $S(b)$ for the nucleus with $A$ nucleons
which is defined as usual 
\beq\label{F16}
S(b)\,=\,\frac{3}{4\pi}\,\frac{1}{R_{A}^{3}\,+\,a^2\,\pi^2\,R_{A}}\,
\int_{-\infty}^{\infty}\,\frac{dz}{1\,+
\,\exp\Le\,\frac{-R_{A}+\sqrt{b^2\,+\,z^2}}{a}\Ra}\,
\eeq
with the parameters
\beq\label{F17}
R_{A}\,=\,5.7\,A^{1/3}\,GeV^{-1}\,,\,\,a\,=\,2.725\,GeV^{-1}\,.
\eeq
It must be underlined, that the introduced Wood-Saxon nuclei profile \eq{F16}
is a attempt of the generalized description of the all nuclei 
density profiles in one
formula. In reality the two parameter Fermi model expression  
for nuclei density \eq{F16} is mostly applied
for heavy nuclei and it is not so correct for the light ones, see \cite{DeJag}.
Therefore, the precise consideration of  
the processes with the light nuclei involved  
need a introduction of the different from the  expression \eq{F16} nuclei density 
profiles and we are not consider this particular 
task in the paper.  

  The $C$ parameter in \eq{F15}, as well as the value of $\alpha_s$, value
of initial rapidity $y_0$ ($x_0$) and masses of three light quarks 
we take the same for the both cases of
DIS processes on the proton and nuclei, see Table\ref{Param}.
\begin{table}[hptb]
\begin{center}
\begin{tabular}{|c|c|c|c|c|}
\hline
\, & \,& \,& \, &\,\\
$y_0\,(x_0)\,$ &  $R^{2}_{p}\,\,(GeV^{-2})$ & $C$ &
$\alpha_{s}$ & $m_{q}^{2}\,\,(GeV^{2})$ \\
\, & \,& \,& \, &\,\\
\hline
\, & \,& \,& \, &\,\\
$3.1\,\,(0.045) $ & 7.9 & 0.0855 & 0.108 & 0.008 \\
\, & \,& \,& \, &\,\\
\hline
\end{tabular}
\caption{\it The parameters of the model.}
\label{Param}
\end{center}
\end{table}
There is a question, is this correct approach to use the same values of the parameters for both processes 
with the different targets. In principle, 
due the larger parton densities in the DIS processes on the nuclei, 
we could expect the differences in 
the values of  $\alpha_{s}$ and values
of $y_0$ for these two  DIS processes.
Nevertheless, we must remember that we perform the calculation in the LO
and running coupling effects are not directly included in our calculation scheme.
We also will see, that a difference in the saturation momenta of the proton and 
different nuclei is not so large, 
and , therefore, this difference could not set up the 
large differences in value of $\alpha_{s}$ for both processes in the LO calculations. 
The justification of the use of the same values of $\alpha_{s}$ and 
the correctness of used parameters will be 
shown in the next section, where surprising coincide of our calculations 
of the integrated gluon density function with the existing parameterizations of the same 
function will be demonstrated. 
The same values of the $C$ parameter and value of initial rapidity, from which the 
evolution equation is valid, may be explained by the same observations.
As it seems, 
these values are more or less universal in the leading order of calculations
and shows 
the physical boundary for the small x evolution independently from the
considered target in DIS process. 

 Last remark which concerns the presented calculations is about the 
kinematic range of the considered parameters. How it was shown in \cite{Serg1},
in the present framework the $F_2$ HERA data
at $Q^{2}_{0}\,<\,1\,\,GeV^{2}$ could not be described. Therefore, in our calculations we
limit the kinematic range of the processes 
by the $x\,<\,x_0$ and $Q^{2}\,>\,Q^{2}_{0}\,$ constraints. 
Unfortunately it means, that the main bulk of the data of the DIS on nuclei,
\cite{Amaud2,Amaud1}, is outside of the range aof the applicability of the model, 
see also remarks above about 
the form of the nuclei density profile. Therefore, in the next section,
we perform the comparison of our results only with the existing 
parameterizations of the integrated gluon density function , without an introducing
of the experimental data fit as it was done in \cite{Serg1}.

\section{Integrated gluon density function}

\begin{figure}[hptb]
\begin{tabular}{ c c}
\psfig{file=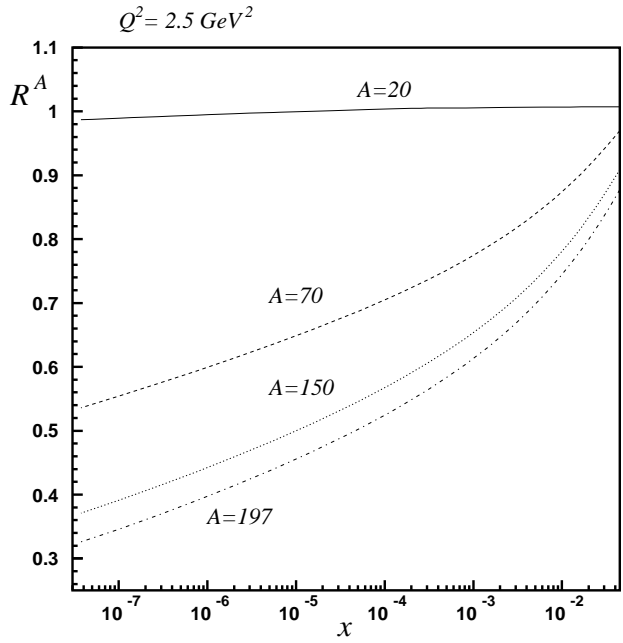,width=90mm} &
\psfig{file=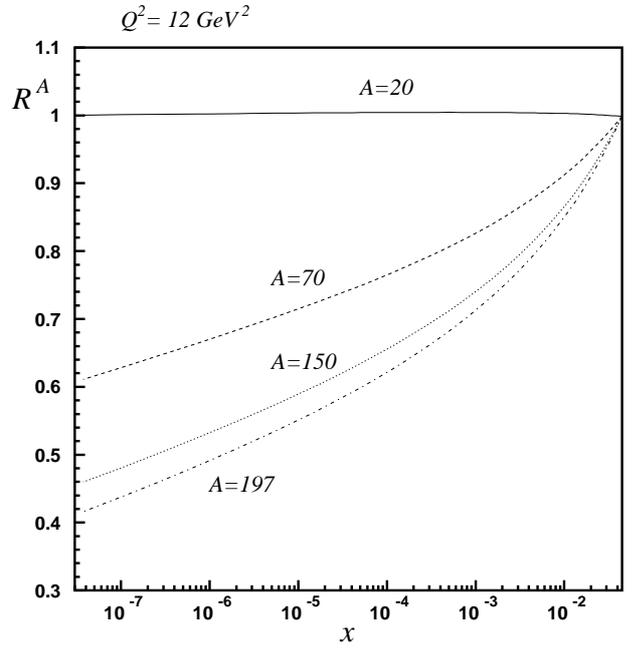,width=90mm}\\
 &  \\
\psfig{file=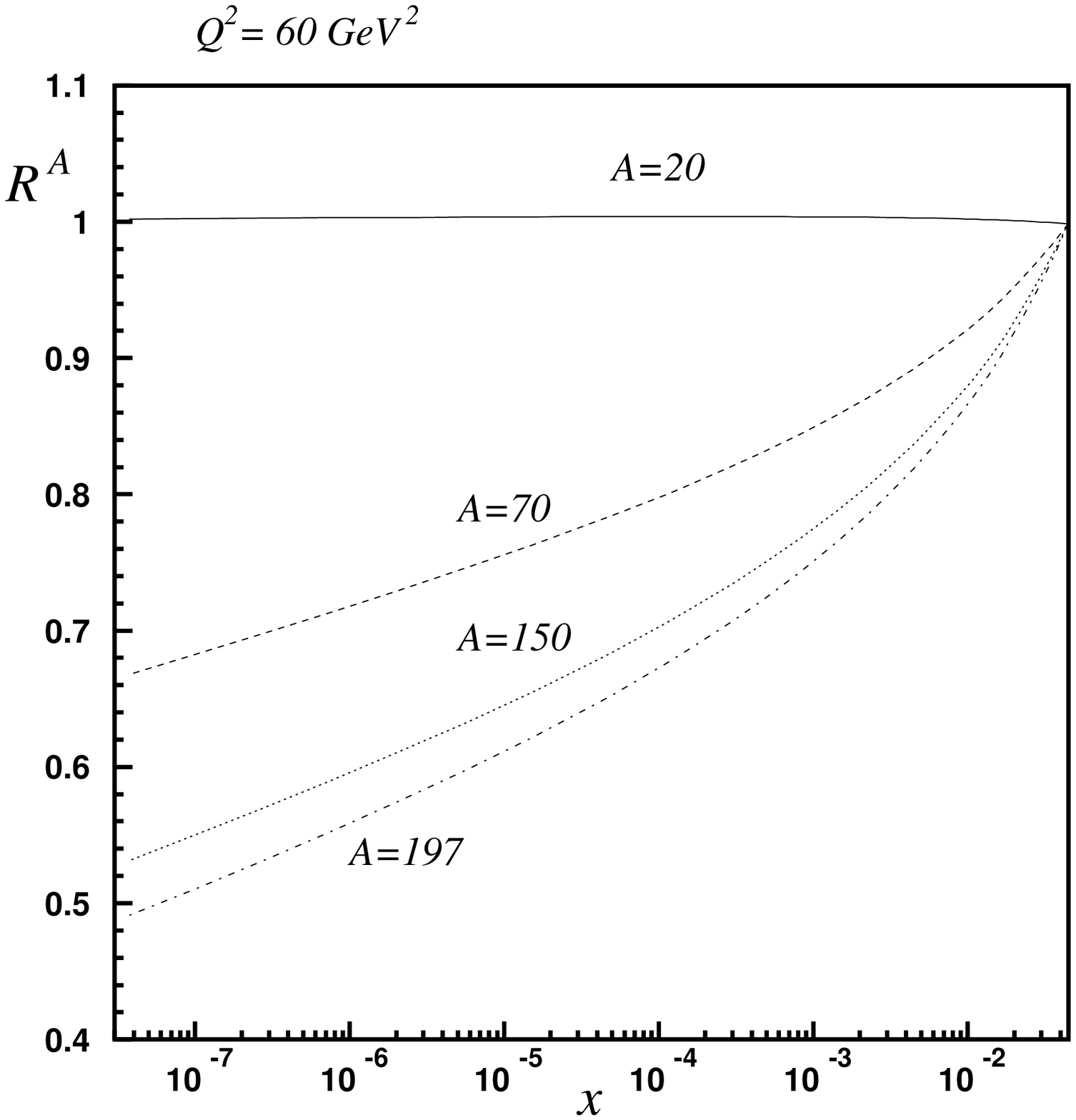,width=90mm} & \,
\psfig{file=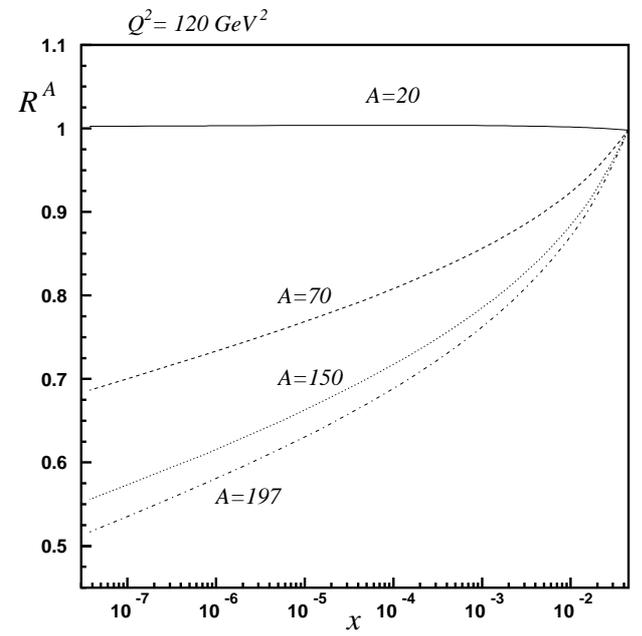,width=90mm}\\
 &  \\ 
\end{tabular}
\caption{\it The ratio $R^{A}(x,Q^2)$  
for different nuclei at different $Q^2$.}
\label{Fig11}
\end{figure}

\begin{figure}[hptb]
\begin{tabular}{ c c}
\psfig{file=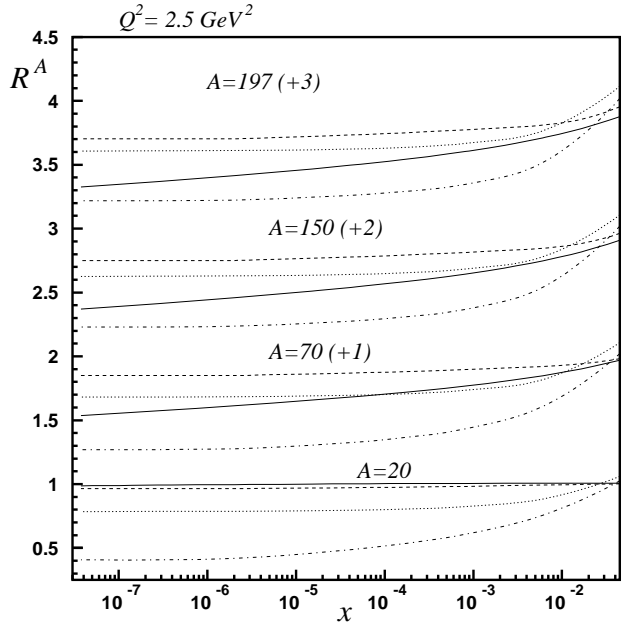,width=90mm} &
\psfig{file=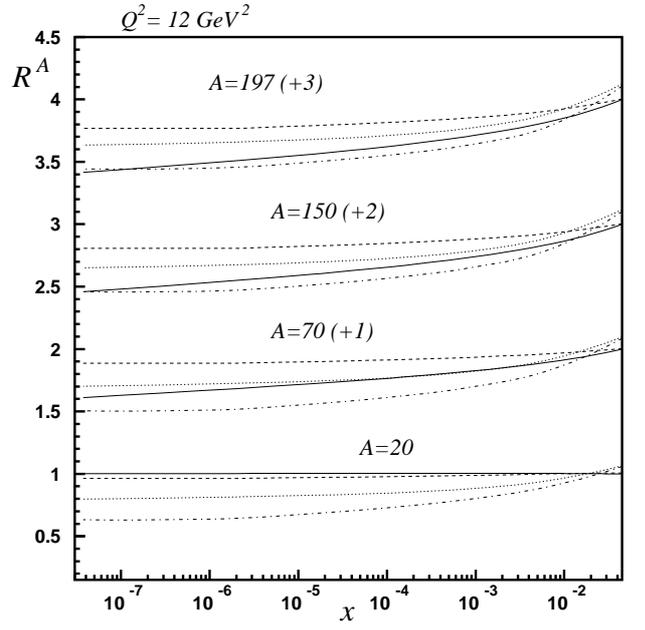,width=90mm}\\
 &  \\
\psfig{file=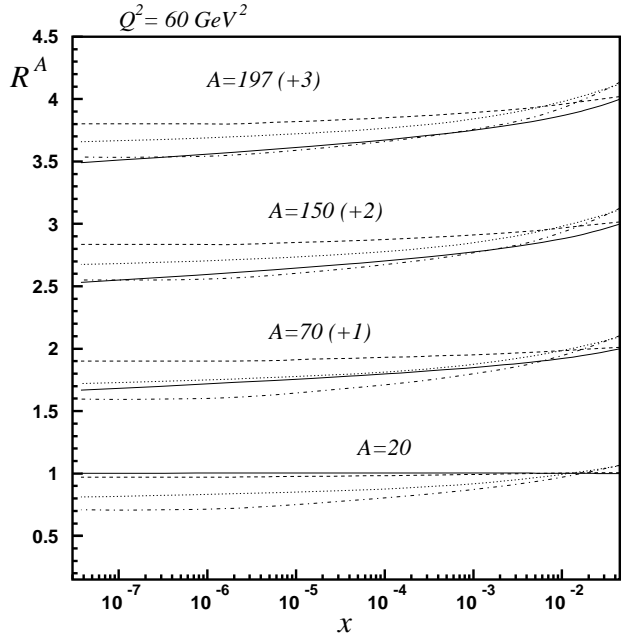,width=90mm} & \,
\psfig{file=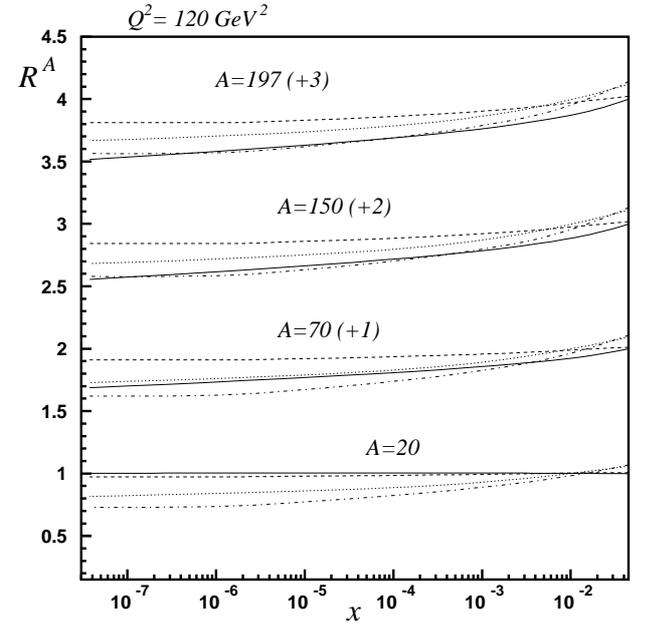,width=90mm}\\
 &  \\ 
\end{tabular}
\caption{\it The ratio $R^{A}(x,Q^2)$ 
of integrated gluon density functions:
the present model results (solid lines), the results from
\cite{Flor} (dashed lines), the results from
\cite{Eskola1} (dotted lines lines) and the results from
\cite{Eskola2} (dashed-dotted lines). The numbers in brackets
show the added number to the real value of $R^{A}$.}
\label{Fig1}
\end{figure}

\begin{figure}[hptb]
\begin{tabular}{ c c}
\psfig{file=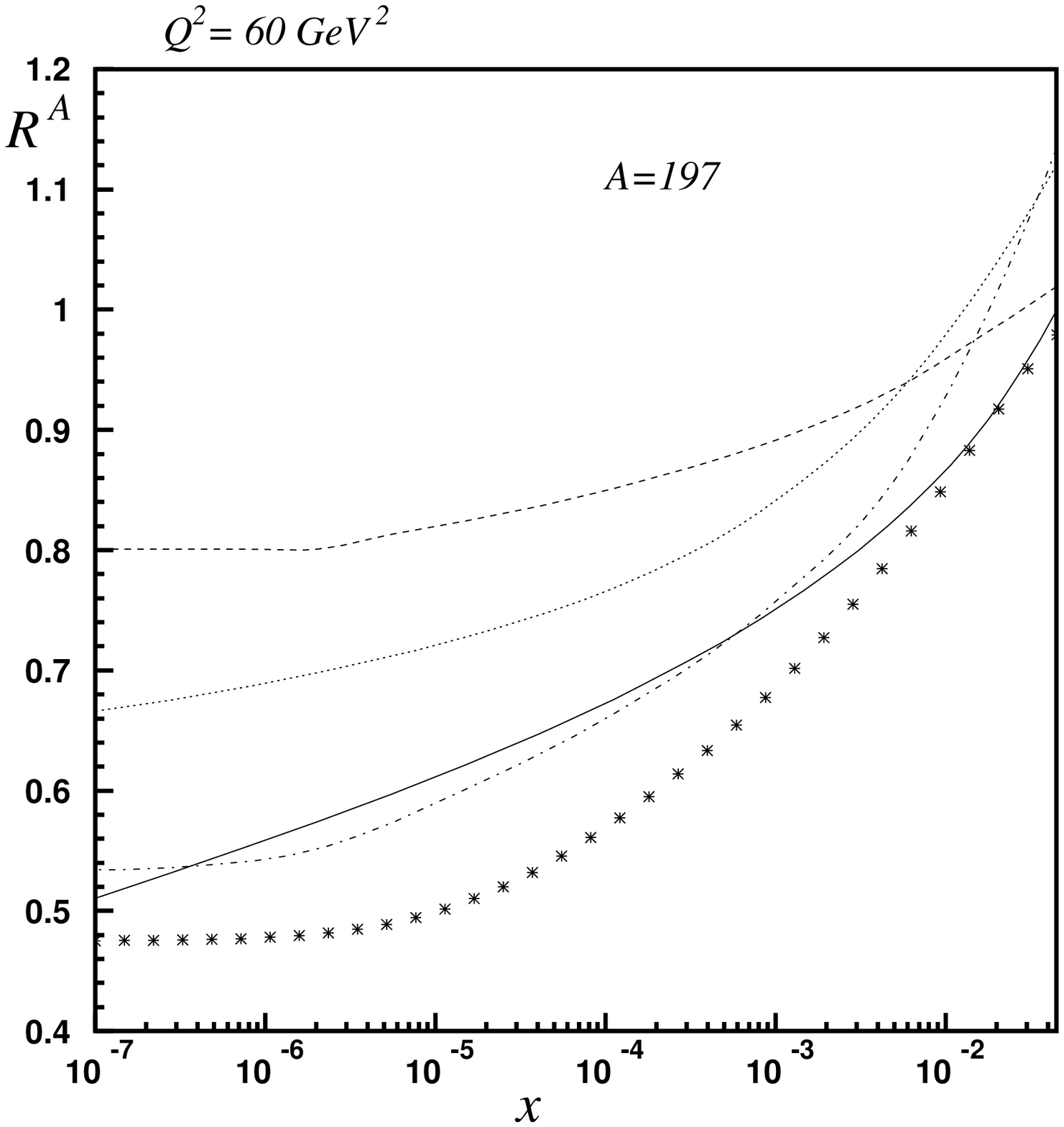,width=90mm} &
\psfig{file=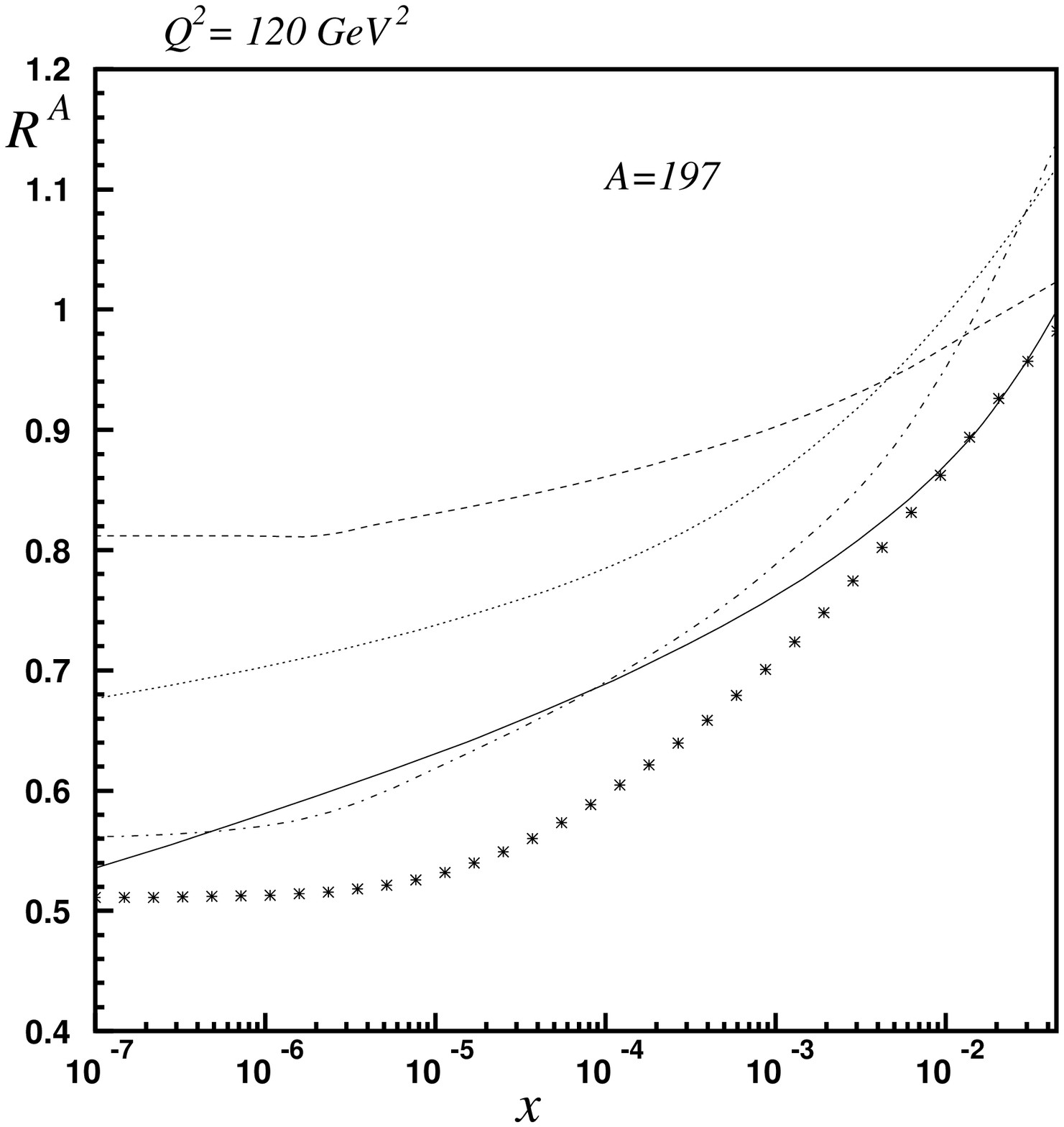,width=90mm}\\
 &  \\
\end{tabular}
\caption{\it The ratio $R^{A}(x,Q^2)$ 
of integrated gluon density functions for the DIS on the gold
at $Q^{2}=60,120\,\,\,GeV^{-2}$:
the present model results (solid lines), the results from
\cite{Flor} (dashed lines), the results from
\cite{Eskola1} (dotted lines lines), the results from
\cite{Eskola2} (dashed-dotted lines) and the results from
\cite{Tywon} (stars).}
\label{Fig111}
\end{figure}

 In \cite{Serg1} was mentioned, that the definition of the integrated gluon density 
function $xG(x,Q^2)$ through the $\tilde{f}(\ytau,k^2,b)$ function of  Eq.\ref{F4}
has ambiguities related with the coupling constant $\alpha_{s}$. Indeed, by
definition
\beq\label{Gd1}
xG(x,Q^2)\,=\,\int\,d^2\,b\,\int^{Q^2}\,\frac{d\,k^2}{k^2}\,
f(x,k^2,b)\,
\eeq
or, in terms of $\tilde{f}(\ytau,k^2,b)$ function
\beq\label{Gd2}
xG(x,Q^2)\,=\,\int\,d^2\,b\,\int^{Q^2}\,\frac{d\,k^2}{k^2}\,
\frac{\tilde{f}(x,k^2,b)}{\alpha_s(k^2)}\,.
\eeq
We see from Eq.\ref{Gd2} that correct determination of the $xG(x,Q^2)$
function in terms of the $\tilde{f}(x,k^2,b)$
function must include the integration over the running coupling, whereas
our calculations scheme includes only LO corrections. In LO approximation
the Eq.\ref{Gd2} may be redefined as
\beq\label{Gd3}
xG(x,Q^2)\,=\,\frac{1}{\alpha_s(<Q^{2}_{proton}>)}
\int\,d^2\,b\,\int^{Q^2}\,\frac{d\,k^2}{k^2}\,\tilde{f}(x,k^2,b)\,
\eeq
where
the value $\alpha_s(<Q^{2}_{proton}>)$ could be considered as a some parameter,
which is not necessary the same as in Table\ref{Param}.
Similarly, the integrated  gluon density function for a nucleus
we could write in the following form 
\beq\label{Gd4}
xG_{N}(x,Q^2)\,=\,\frac{1}{\alpha_s(<Q^{2}_{nucleus}>)}
\int\,d^2\,b\,\int^{Q^2}\,\frac{d\,k^2}{k^2}\,\tilde{f}(x,k^2,b)\,,
\eeq
with the $\tilde{f}(\ytau,k^2,b)$ as a solution of corresponding BK equation
for DIS process on the nucleus. Our main assumption, which we could check only 
post priori, is that in LO we could take
\beq\label{Gd5}
\alpha_s(<Q^{2}_{proton}>)\,\approx\,\alpha_s(<Q^{2}_{nucleus}>)
\eeq
and, therefore, that the ratio of the integrated gluon densities of the nucleus
with $A$ nucleons and the proton
\beq\label{Gd6}
R^{A}(x,Q^2)\,=\,\frac{xG_{N}(x,Q^2)}{A\,xG(x,Q^2)}
\eeq
does not depend on values of $\alpha_{s}$. The first check 
of the approximation made is simple. We calculate 
$R^{A}$ for the following nuclei: 
gold (A=197), neodymium (A=150), zinc (A=70) and neon (A=20)
at 
$$
Q^{2}\,=\,2.5\,,\,12\,,\,60\,,\,120\,\,GeV^{2}\,.
$$
Obtained results, see Fig.\ref{Fig11},  we compare 
with the calculations of the ratio from the 
\cite{Flor,Eskola1,Eskola2,Tywon}, see 
Fig.\ref{Fig1}, Fig.\ref{Fig111}.
As it seems from the Fig.\ref{Fig1}-Fig.\ref{Fig111}, in general our
results for $R^{A}$  are in the range defined by curves obtained by the 
parameterization \cite{Eskola1,Eskola2} for $R^{A}$ ratio, and clearely 
more differ from the \cite{Flor,Tywon} parameterizations of the 
same ratio. Surprisingly, obtained in absolutely different framework
our results somehow interpolate between \cite{Eskola1,Eskola2}
parameterization of the ratio and stay in the "window" defined 
at low x by "extremal" parameterizations \cite{Flor,Tywon}.
The
closeness of all curves at the initial point of small x evolution, $(x=0.045)$,
shows that we indeed matched the small x evolution of BK equation
with the DGLAP framework of  \cite{Flor,Eskola1,Eskola2,Tywon} in this point.
Therefore, this coincidence between the curves from different calculation frameworks 
indeed justifies the form of used initial conditions for the BK equation
and the assumption Eq.\ref{Gd5}. It is also clear, that
all parameterizations  \cite{Flor,Eskola1,Eskola2,Tywon} are 
based on the low energy data, whereas
the high energy parts of the curves are the extrapolation of the established
formulae in the region of small x. This is a explanation for the large differences 
between the curves from different parameterization in the region of small x in the  
Fig.\ref{Fig1}. 

\begin{figure}[hptb]
\begin{tabular}{ c c}
\psfig{file=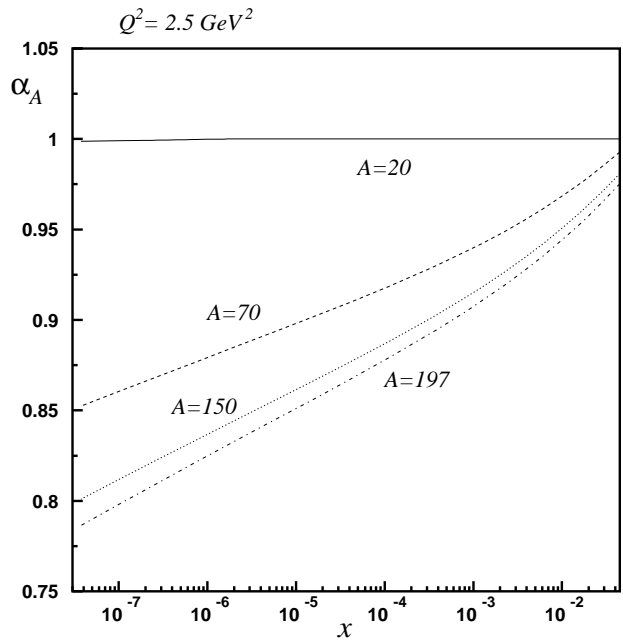,width=90mm} &
\psfig{file=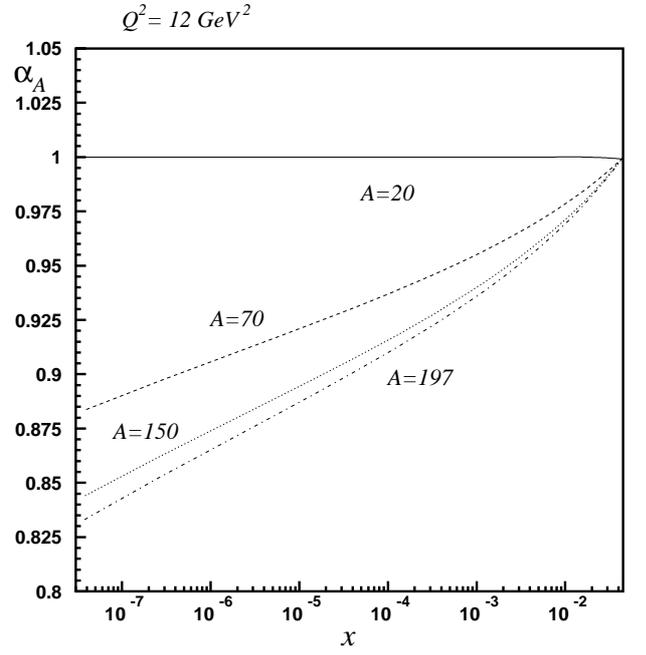,width=90mm}\\
 &  \\
\psfig{file=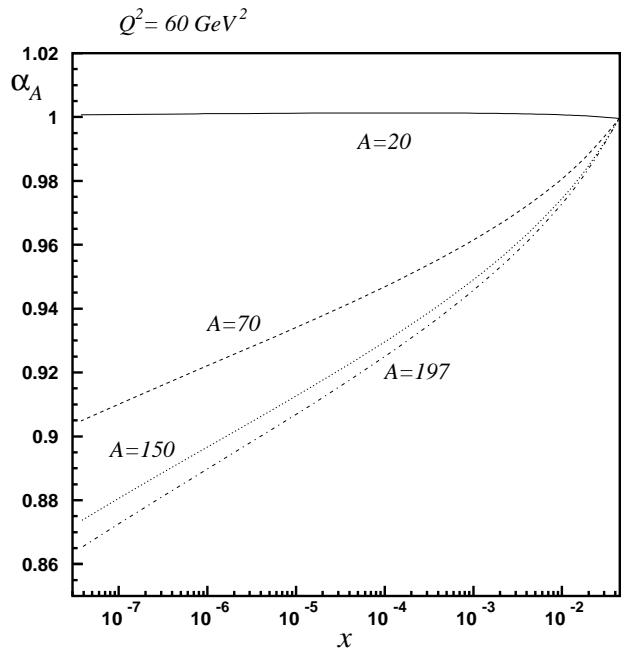,width=90mm} & \,
\psfig{file=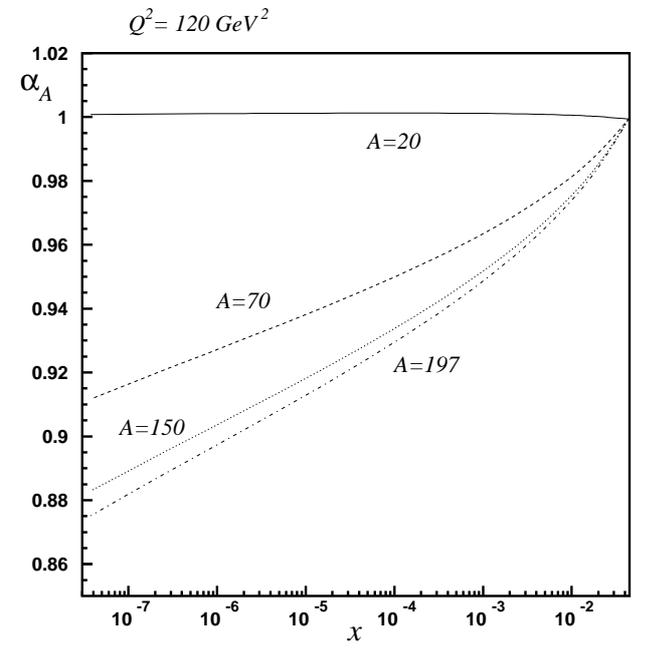,width=90mm}\\
 &  \\ 
\end{tabular}
\caption{\it The $\alpha_{A}$ coefficient of Eq.\ref{Gd7}
for different nuclei at different $Q^2$.}
\label{Fig2}
\end{figure}

Another question is about the parameterization of the nuclei
gluon density $xG_{N}(x,Q^2)$ function in the form 
\beq\label{Gd7} 
xG_{N}(x,Q^2)\,=\,A^{\alpha_{A}}\,xG(x,Q^2)\,
\eeq
where is the coefficient $\alpha_{A}$ (do not be confused with
the coupling constant $\alpha_{s}$) determines the "power"
of the shadowing for each nucleus at different $Q^2$.
In spite of the parameterization based on the data fitting,
in our approach this coefficient is calculable, 
see also \cite{Lev11} for the similar calculations. The
results of the calculation of the 
$\alpha_{A}$ coefficient is presented in the Fig.\ref{Fig2}.
From Fig.\ref{Fig2} it is clear, that the obtained shadowing is weaker than usually
obtained in the framework of BK equation, see again paper 
\cite{Lev11} for example.

\section{$F_2$ structure function}

\begin{figure}[hptb]
\begin{tabular}{ c c}
\psfig{file=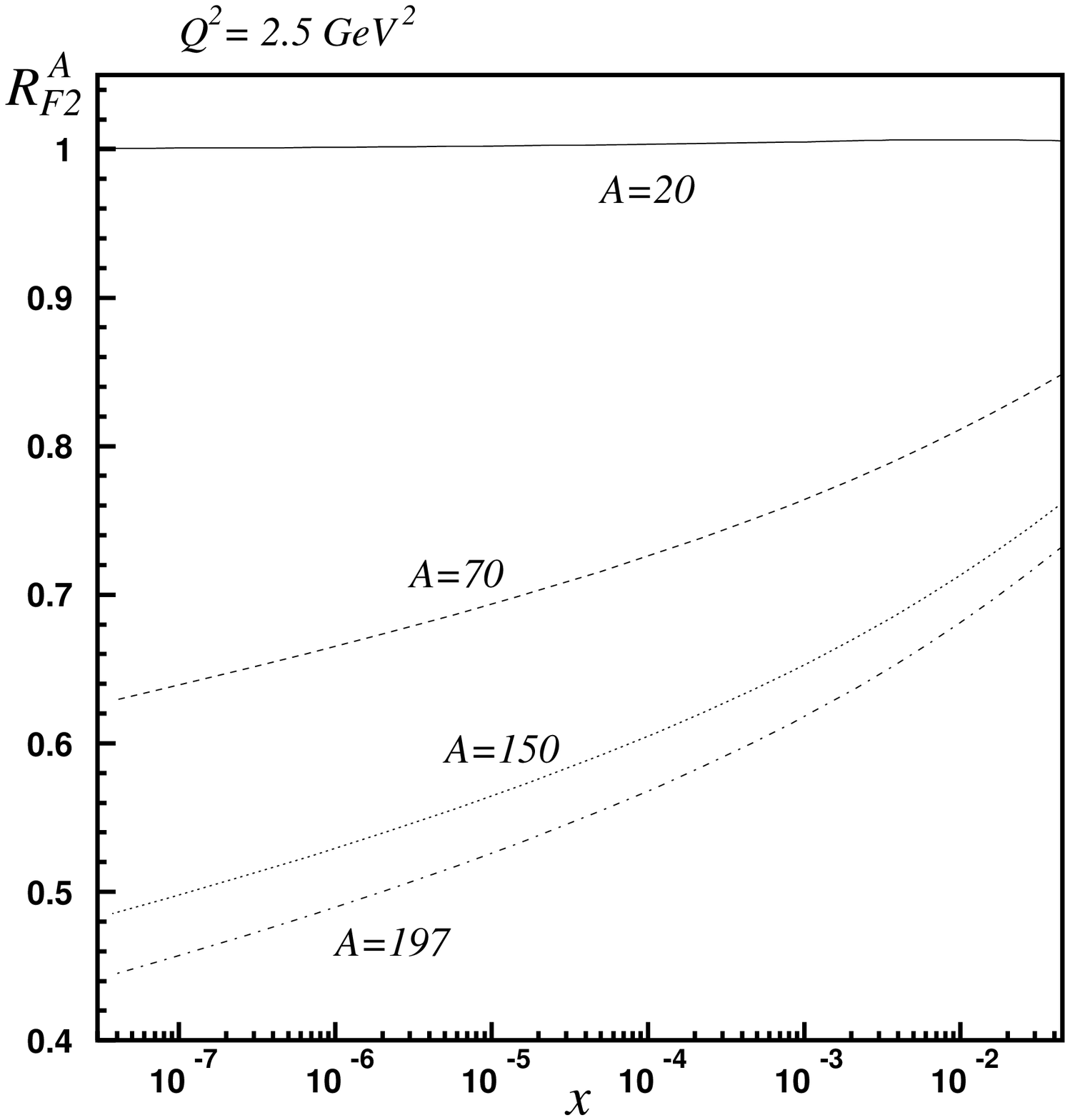,width=90mm} &
\psfig{file=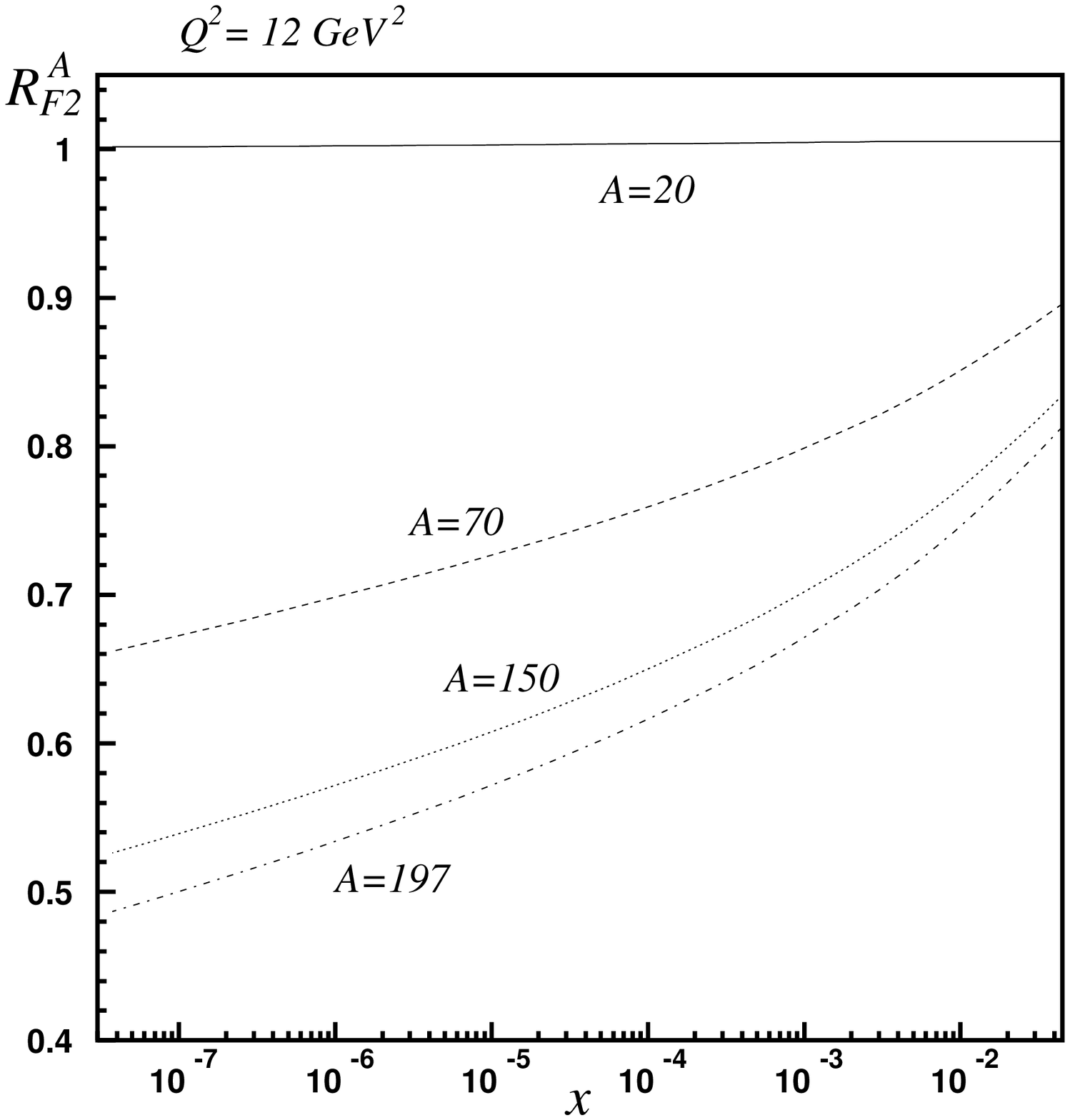,width=90mm}\\
 &  \\
\psfig{file=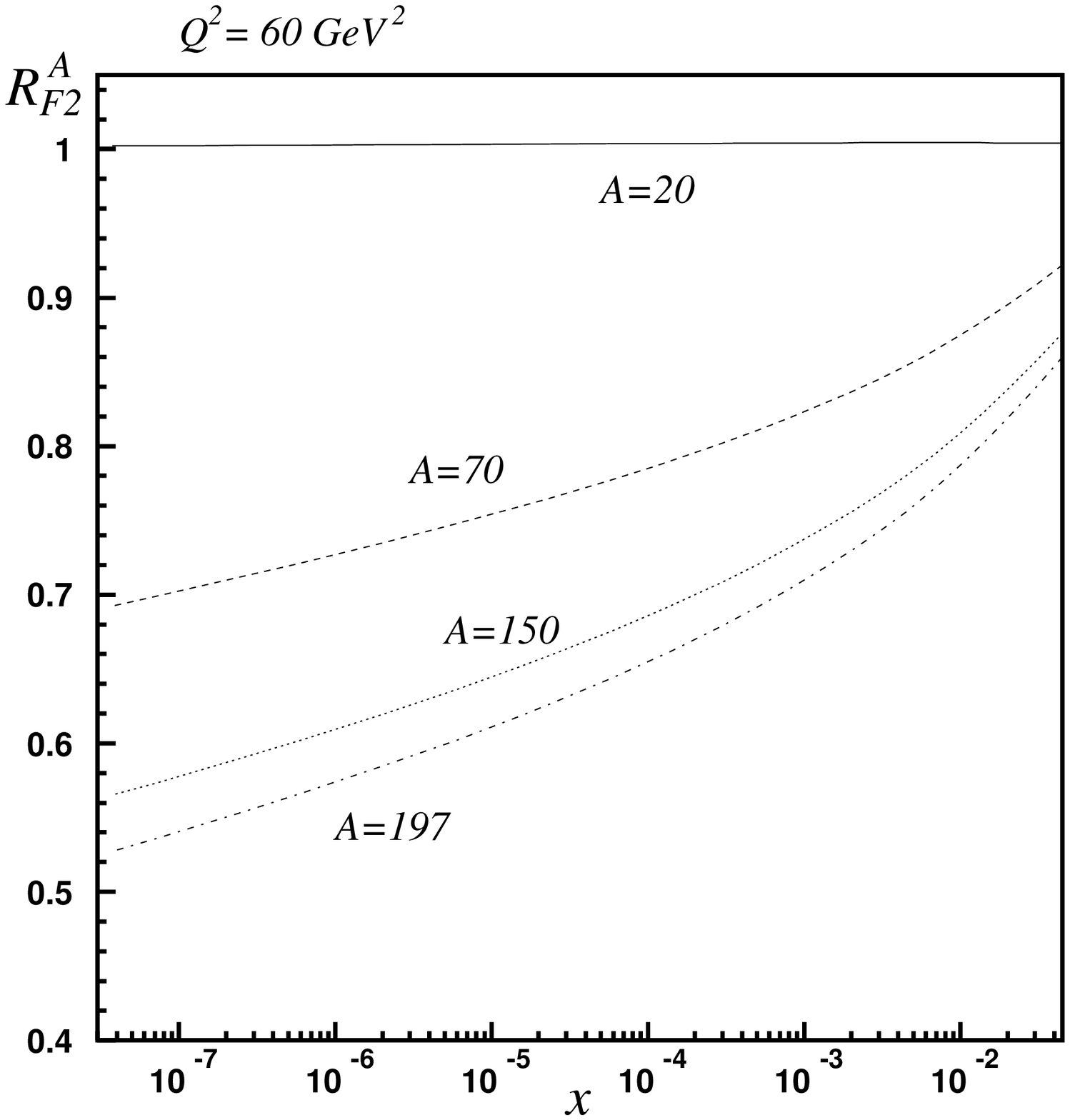,width=90mm} & \,
\psfig{file=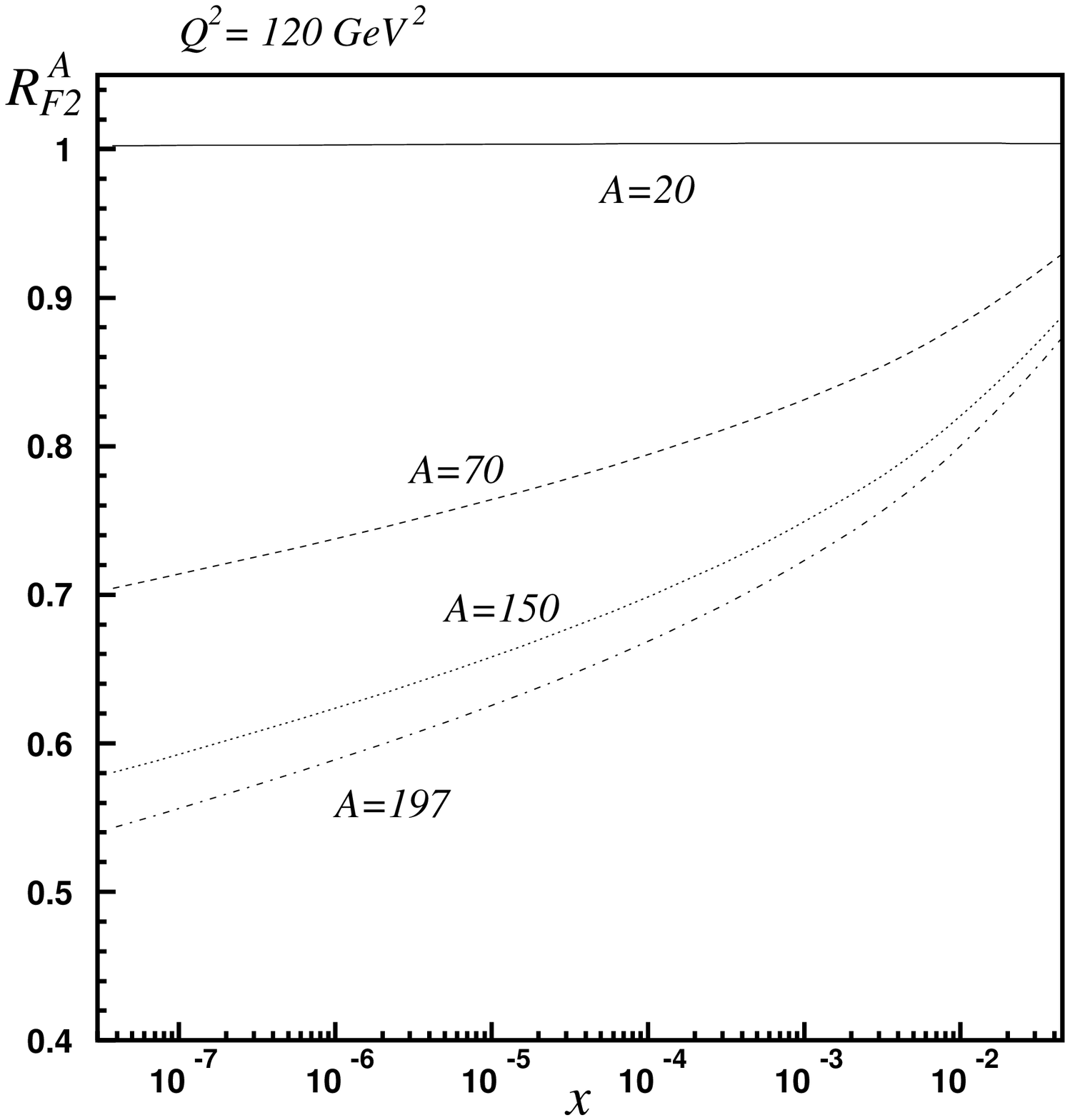,width=90mm}\\
 &  \\ 
\end{tabular}
\caption{\it The $R^{A}_{F2}$ ration for different nuclei at different $Q^2$.}
\label{Fig3}
\end{figure}

\begin{figure}[hptb]
\begin{tabular}{ c c}
\psfig{file=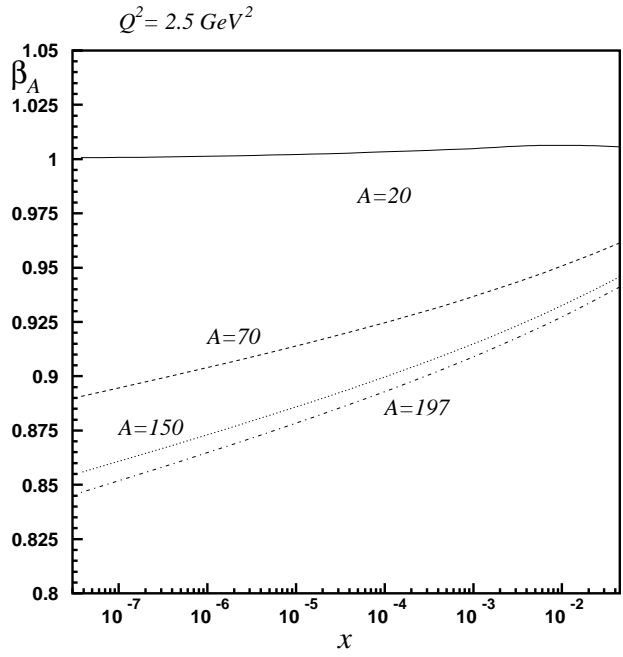,width=90mm} &
\psfig{file=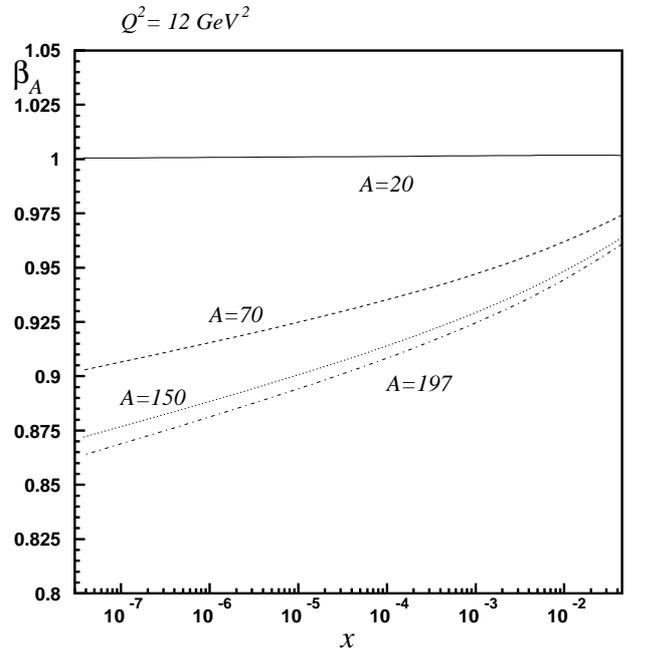,width=90mm}\\
 &  \\
\psfig{file=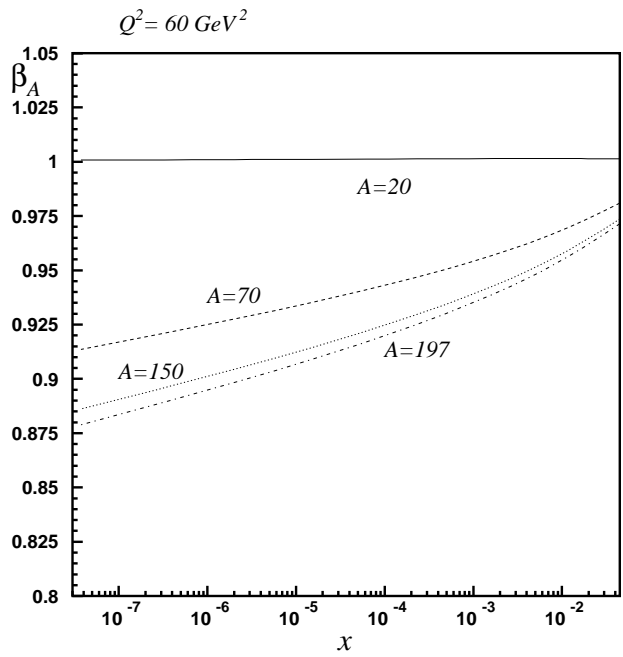,width=90mm} & \,
\psfig{file=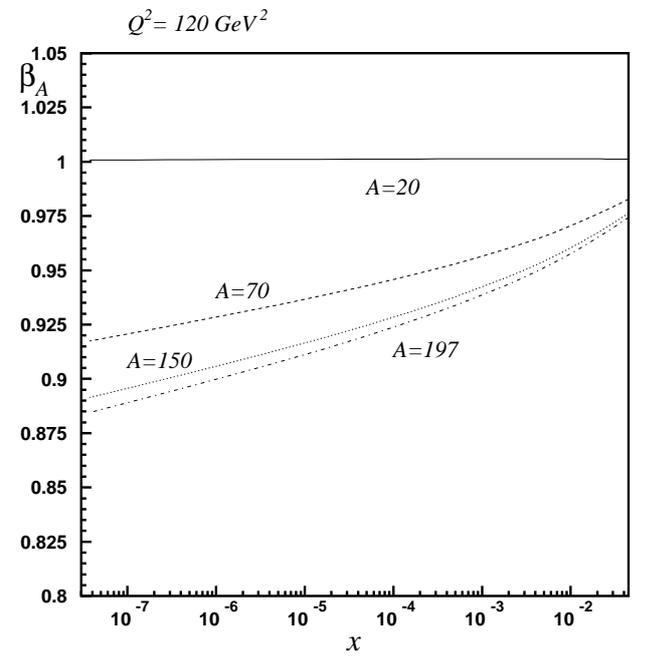,width=90mm}\\
 &  \\ 
\end{tabular}
\caption{\it The $\beta_{A}$ coefficient of Eq.\ref{FS4}
for different nuclei at different $Q^2$.}
\label{Fig4}
\end{figure}

Using usual definition of $F_2$ structure function
\beq\label{FS1} 
F_{2}(x,\,Q^2)\,=\,\frac{Q^2\,\alpha_s}{4\,\pi^2\,\al}\,\int\,d^2\,b\,
\int\,\frac{d^2\,k}{k^4}\,\frac{f(x,\,k^2,b)}{4\,\pi}\,
\Le\,\Phi_{\T}(k,m^{2}_{q})\,+\,\Phi_{L}(k,m^{2}_{q})\,\Ra\,
\eeq
we obtain the same expression in terms of $\tilde{f}(\ytau,k^2,b)$
unintegrated gluon density function
\beq\label{FS2}
F_{2}(x,\,Q^2)\,=\,\frac{Q^2}{4\,\pi^2\,\al}\,\int\,d^2\,b\,
\int\,\frac{d^2\,k}{k^4}\,\frac{\tilde{f}(x,\,k^2,b)}{4\,\pi}\,
\Le\,\Phi_{\T}(k,m^{2}_{q})\,+\,\Phi_{L}(k,m^{2}_{q})\,\Ra\,.
\eeq
The expressions for the impact factors in Eq.\ref{FS1}-Eq.\ref{FS2} are
given in Eq.\ref{F6}-Eq.\ref{F7}. 
There the $\alpha_s$ coupling constant is excluded from the expression comparing 
to the usual definition of the impact factors, that allows to write Eq.\ref{FS2}
in the way where formally $\alpha_s$ is not appeared in the expression.
As in the previous case of integrated gluon density function, the main object of our interest 
is a ratio of the nucleus and proton structure functions
\beq\label{FS3} 
R^{A}_{F2}=\frac{F_{2N}(x,\,Q^2)}{A\,F_2(x,\,Q^2)}\,.
\eeq
The results for this ratio are presented in the  Fig.\ref{Fig3}.
Comparing the result of the Fig.\ref{Fig11} with the results presented in the 
Fig.\ref{Fig3} it is easy to see, that in contrary to the $xG_{N}(x,\,Q^2)$
function the structure function $F_{2N}(x,\,Q^2)$ does not proportional
to the $A\,F_{2}(x,\,Q^2)$even at high x. Introducing
the following parameterization
\beq\label{FS4} 
F_{2N}(x,Q^2)\,=\,A^{\beta_{A}}\,F_{2}(x,Q^2)\,
\eeq
we obtain for the coefficient $\beta_{A}$ results which presented in the 
Fig.~\ref{Fig4}.
As it seems from the Fig.\ref{Fig4} the coefficient $\beta_{A}$
is less sensitive to the details of the process, i.e. to the values of $Q^2$
and type of the nucleus than $\alpha_{A}$ coefficient from   Fig.\ref{Fig3}.
It is also interesting to note , that even at high values of x
the coefficient $\beta_{A}$ does not equal to one, in contrary to the 
$\alpha_{A}$.

\section{Anomalous dimension}

\begin{figure}
\begin{center}
\psfig{file=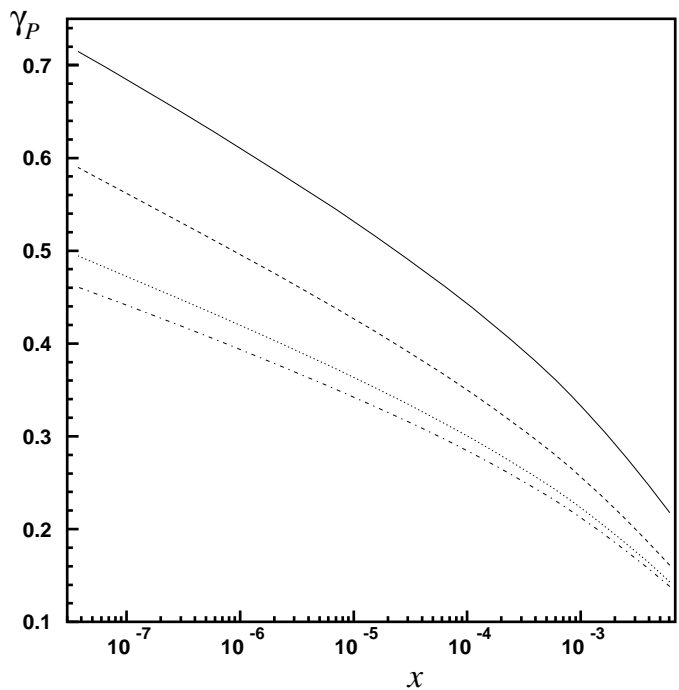,width=100mm} 
\end{center}
\caption{\it Anomalous dimension $\gamma$ for the case of DIS on the proton
at different $Q^2$: $Q^2\,=2.5\,\,GeV^{2}$ (solid line), 
$Q^2\,=12\,\,GeV^{2}$ (dashed line), $Q^2\,=60\,\,GeV^{2}$ (dotted line),
$Q^2\,=120\,\,GeV^{2}$ (dashed-dotted line).}
\label{Fig5}
\end{figure}

\begin{figure}[hptb]
\begin{tabular}{ c c}
\psfig{file=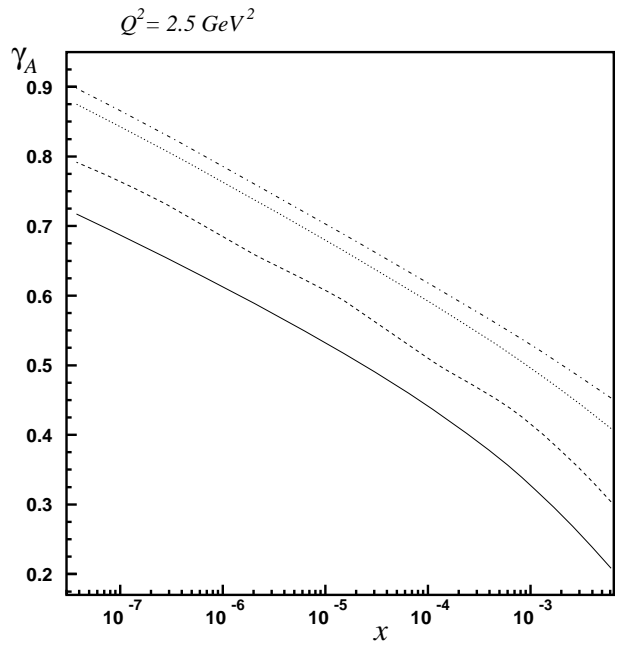,width=90mm} &
\psfig{file=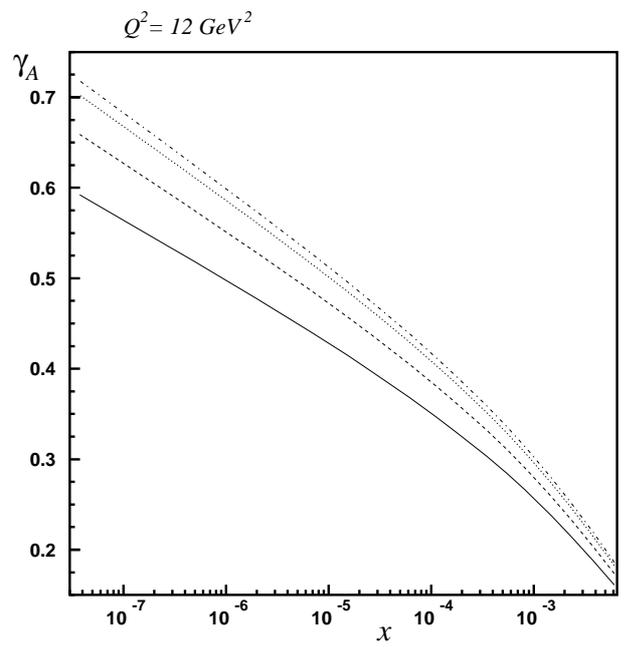,width=90mm}\\
 &  \\
\psfig{file=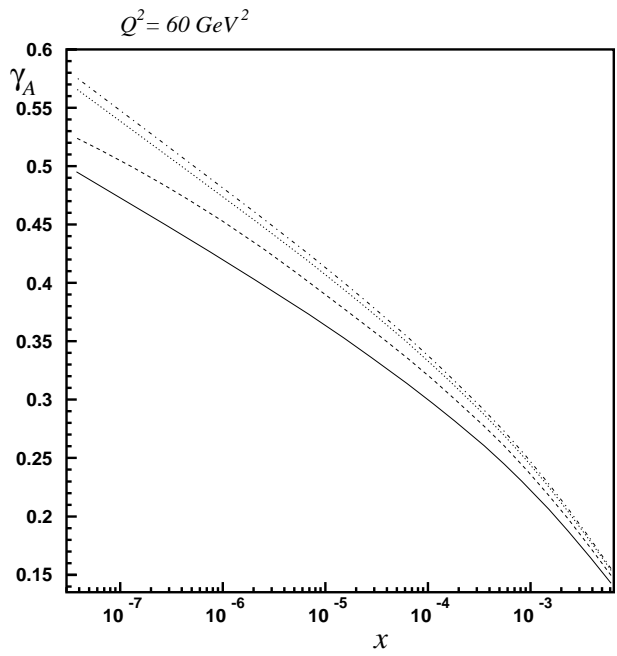,width=90mm} & \,
\psfig{file=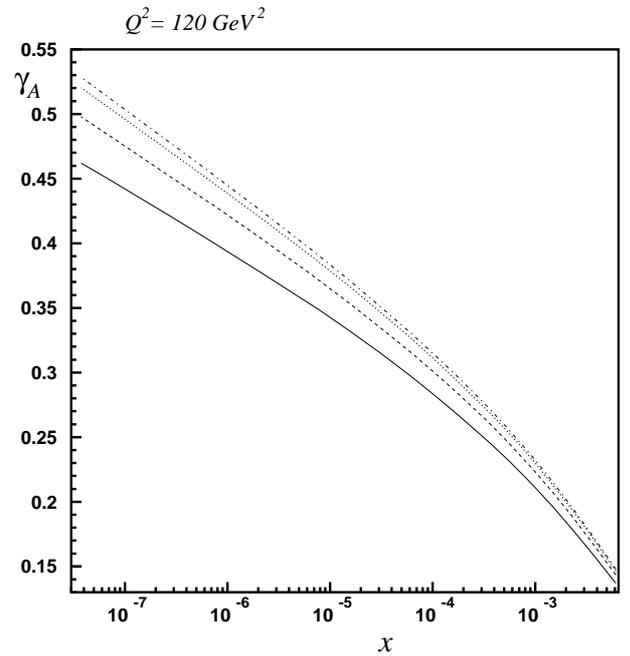,width=90mm}\\
 &  \\ 
\end{tabular}
\caption{\it Anomalous dimension $\gamma$ for the case of DIS on the nuclei:
A=20 (solid line), A=70 (dashed line), A=150 (dotted line), 
A=197 (dashed-dotted line).}
\label{Fig6}
\end{figure}

 Let's consider the definition of the average anomalous
dimension $\gamma$ in DIS process via the integrated gluon density
function 
\beq\label{An1}
xG(x,Q^2)\,\propto\,\Le\,Q^2\,\Ra^{\gamma}\,
\eeq
see \cite{Ayala1}. In this case the calculation of $\gamma$ 
is straightforward
\beq\label{An2}
\gamma\,=\,\frac{\partial\ln\Le\,xG(x,Q^2)\,\Ra}{\partial\ln Q^2}
\eeq
The result for $\gamma$ for the case of DIS process on the proton is represented in 
Fig.\ref{Fig5} and for the case of DIS on the nuclei in the Fig.\ref{Fig6}.
Considering the 
$\gamma_P$ anomalous dimension from the Fig.\ref{Fig5}, it is interesting to note,
that for DIS process on the proton at $Q^2\,>\,60\,\,GeV^{2}$ the value of 
$\gamma_P$ is below the value of BFKL anomalous dimension 
$\gamma_{BFKL}\,\approx\,0.5$ at whole range of x. 
At small values of  $Q^2$ the $\gamma_P$ value is about BFKL 0.5 already at  
$x\,\propto\,\,10^{-4}\,-10^{-5}\,$, that one more time underline the importance
of rescattering correction (shadowing corrections) in this kinematic region.
From the result of the calculations of anomalous dimension for the different nuclei
$\gamma_A$ in Fig.\ref{Fig6} we see, that in contrary to the calculations 
of \cite{Lev11} the  $\gamma_A$ shows clear dependence on the atomic number A at all values of $Q^2$
at small $x\,<\,10^{-4}\,$. Decrease of the value
of $Q^2$ in the DIS process leads to the increase of the value of $\gamma_A$,
that indicates the increasing of value of the shadowing corrections in 
the process.

\section{Saturation momenta} 

 There are different definitions of the saturation momenta,
which are used through the literature about the subject.
For example, in papers \cite{Lev11} the saturation momenta in DIS process
was defined with the help of a packing factor
$\kappa_{p}(Q^2,x)$. As a saturation momenta there was considered a momenta
where
\beq\label{Sm1}
\kappa_{p}(Q^{2}_{S},x)\,=\,1/2\,
\eeq
at fixed x and impact parameter (if, of course, the impact parameter
is introduced in definition of $\kappa_{p}(Q^2,x)$). In our paper
we use a different definition of saturation momenta, borrowed
from \cite{Armesto1}. Following the definition 
of \cite{Armesto1} we define a saturation momenta
as a momenta where a maximum of the unintegrated gluon density function is 
reached at fixed impact parameter and fixed x:
\beq\label{Sm2}
Q_{S}(b,x)\,:\,\frac{\tilde{f}(x,Q_{S},b)}{k^2}\,>\,
\frac{\tilde{f}(x,k,b)}{k^2}\,\,for\,\,any\,\,k\,:\,k_{min}\,<\,k\,<\,k_{max}\,\,,
\eeq
where $k_{min}$  and $k_{max}$ are correspondingly
minimum and maximum momenta used in numerical calculations.
This $Q^{2}_{S}(b)$ definitely depends on impact parameter and, in fact,
may be used in order to introduce the impact parameter dependence in 
the scaling solution of usual BK equation. 
In the next subsections we will use the definition Eq.\ref{Sm2}
for the calculations of saturation momenta of the 
proton and different nuclei.

\subsection{Saturation momenta for DIS process on the proton}

Plot of the saturation momenta of the proton,
defined through the Eq.\ref{Sm2}, are presented in the 
Fig.9-Fig.\ref{Fig7-b} .
\begin{figure}[hptb]
\begin{center}
\psfig{file=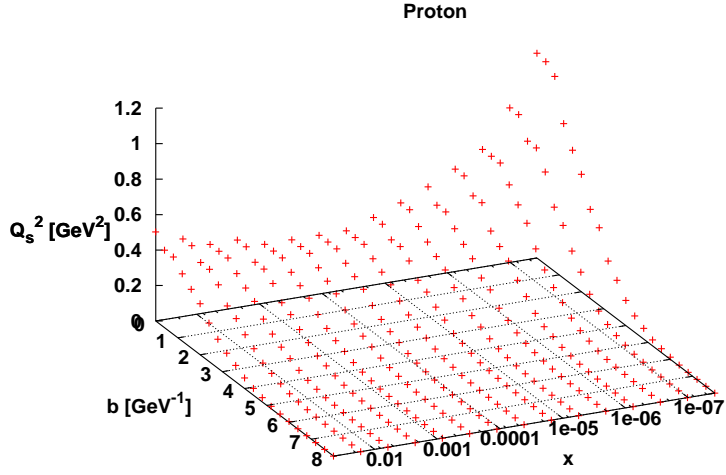,width=105mm} 
\caption{\it Saturation momenta of the proton as a function of x and
impact parameter.}
\end{center}
\label{Fig7}
\end{figure}
Considering the saturation momenta at fixed impact parameters,
we could write a very simple expression for the approximate parameterization 
of the saturation momenta of the proton
\beq\label{SmP1}
Q_{S}^{2}(b,x)\,=\,Q_{S0}^{2}(b)\,+\,Q_{S1}^{2}(b)\,\Le\,
\frac{x_0}{x}\,\Ra^{d(b)}\,,
\eeq
where the coefficients $\,Q_{S0}^{2}(b)\,,\,Q_{S1}^{2}(b)\,$ 
and $\,d(b)\,$ in this parameterization could be extracted from the 
Fig.9 data. The fitting procedure for the
data in the range of x such that 
$\,x\,=\,1.66\,10^{-2}-3.75\,10^{-8}\,$ gives the following
values of coefficients
\beq\label{SmP2}
Q_{S}^{2}(b,x)\,=\,F_{S}(x)\,S(b)\,=\,
\Le\,8.69\,+\,12.65\,\Le\,\frac{10^{-7}}{x}\,\Ra^{0.46}\,\Ra\,
\frac{e^{-b^2/R_{P}^2}}{\pi\,R_{p}^{2}}\,
\,\,GeV^{2}\,,
\eeq
with the coefficient $\,d\,$ which does
not dependent on impact parameter and with the 
proton radius  from the Table~\ref{Param}.
Comparing the expression of the saturation momenta Eq.\ref{SmP2}
with the coefficient $C$ from the expression Eq.\ref{F11}, we see, that
the expression Eq.\ref{SmP2} gives $\,C\,=\,0.11\,GeV^{2}$ instead
$\,C\,=\,0.0855\,GeV^{2}$ in initial conditions Eq.\ref{F11}. The difference
between these two values is due the "averaging" procedure used in fitting the 
data. The expression Eq.\ref{SmP2} is the result of fitting of many points,
and there is not necessary that the resulting curve will cross precisely
the initial point in the fitted data.  
Obtained value of the coefficient $d=0.46$ is close to the results of
\cite{Armesto1}. We obtained $c=2.2$ instead $c=2.06$ in the terms of the 
paper \cite{Armesto1}. 
\begin{figure}[hptb]
\begin{tabular}{ c c}
\psfig{file=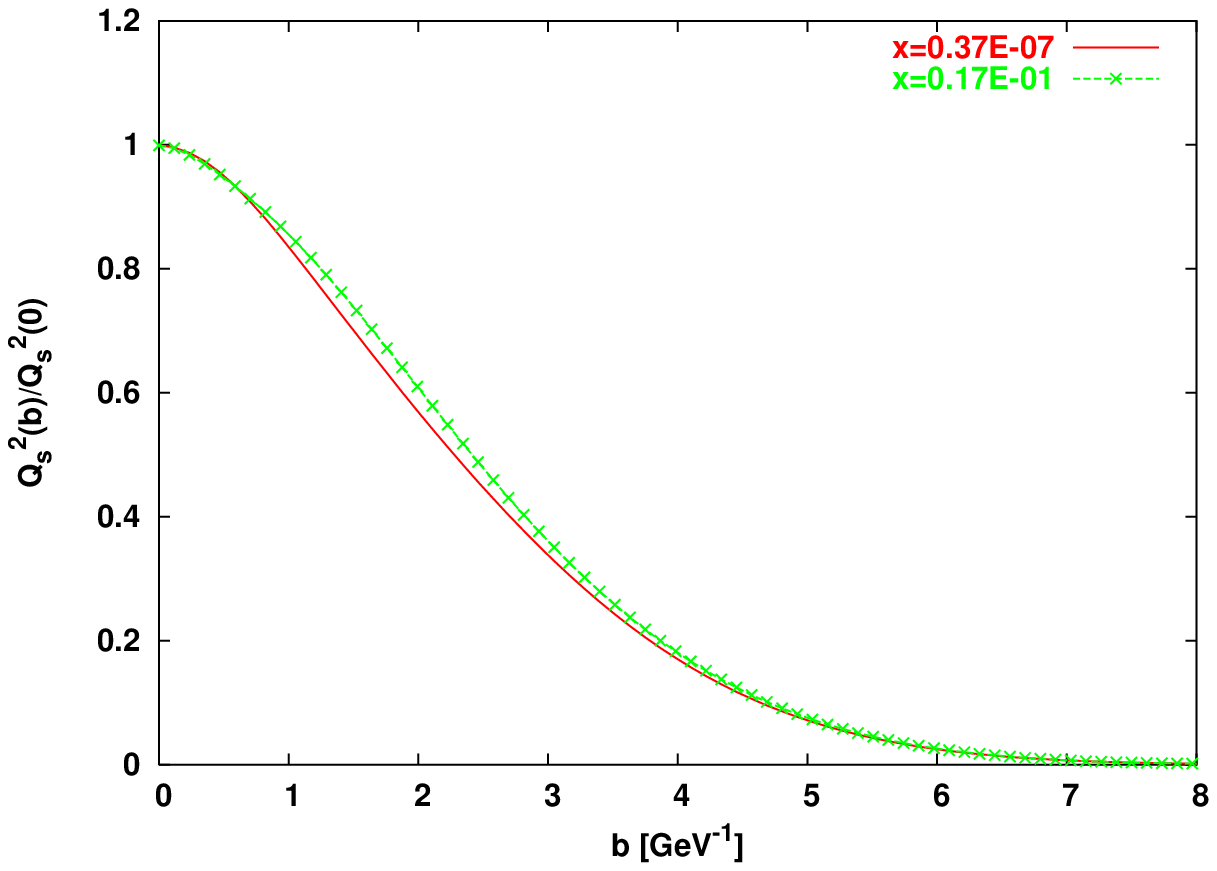,width=80mm} &
\psfig{file=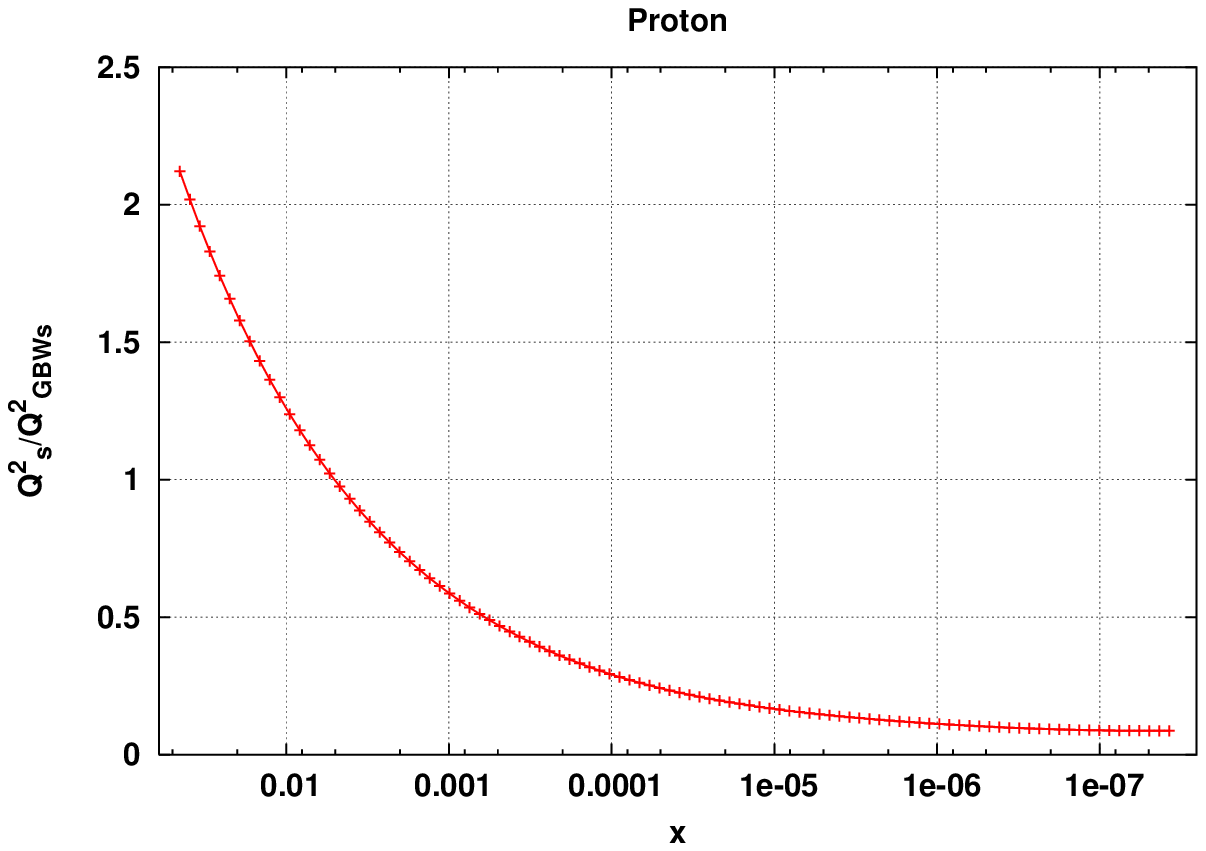,width=86.5mm}\\
 &  \\
\end{tabular}
\caption{\it Saturation momenta as a function of impact parameter at fixed values of
x in the left plot and the comparison of the obtained saturation momenta at $b=0$
with the GBW saturation momenta from the \cite{GolecBiernat1} in the right plot of the picture.}
\label{Fig7-b}
\end{figure}
From the \eq{SmP2}  and Fig.\ref{Fig7-b} we see, that our answer 
for the saturation momenta is different from the obtained in the
GBW model \cite{GolecBiernat1}. We will discuss these differences
in the conclusion of the paper.

\begin{figure}
\begin{center}
\psfig{file=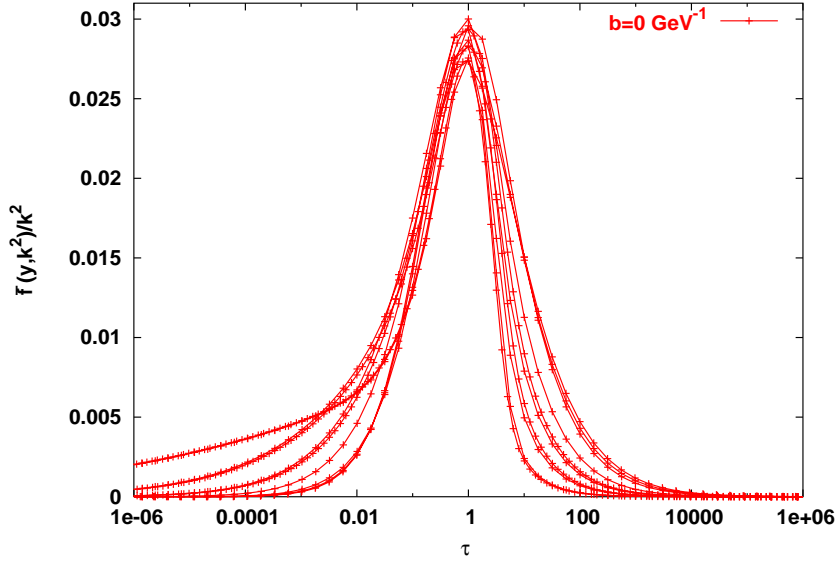,width=110mm} 
\end{center}
\caption{\it Unintegrated gluon density function 
$\tilde{f}(x,k^2,b)\,/\,k^2$ as a function of scaling variable $\tau\,$ 
at $b=0$ and different $x\,=6.1\,10^{-3}\,-
\,3.8\,10^{-8}\,$ (curves from up to down in the left half of the plot) 
for the case of DIS on the proton.}
\label{Fig7-a}
\end{figure}

\begin{figure}[hptb]
\begin{tabular}{ c c}
\psfig{file=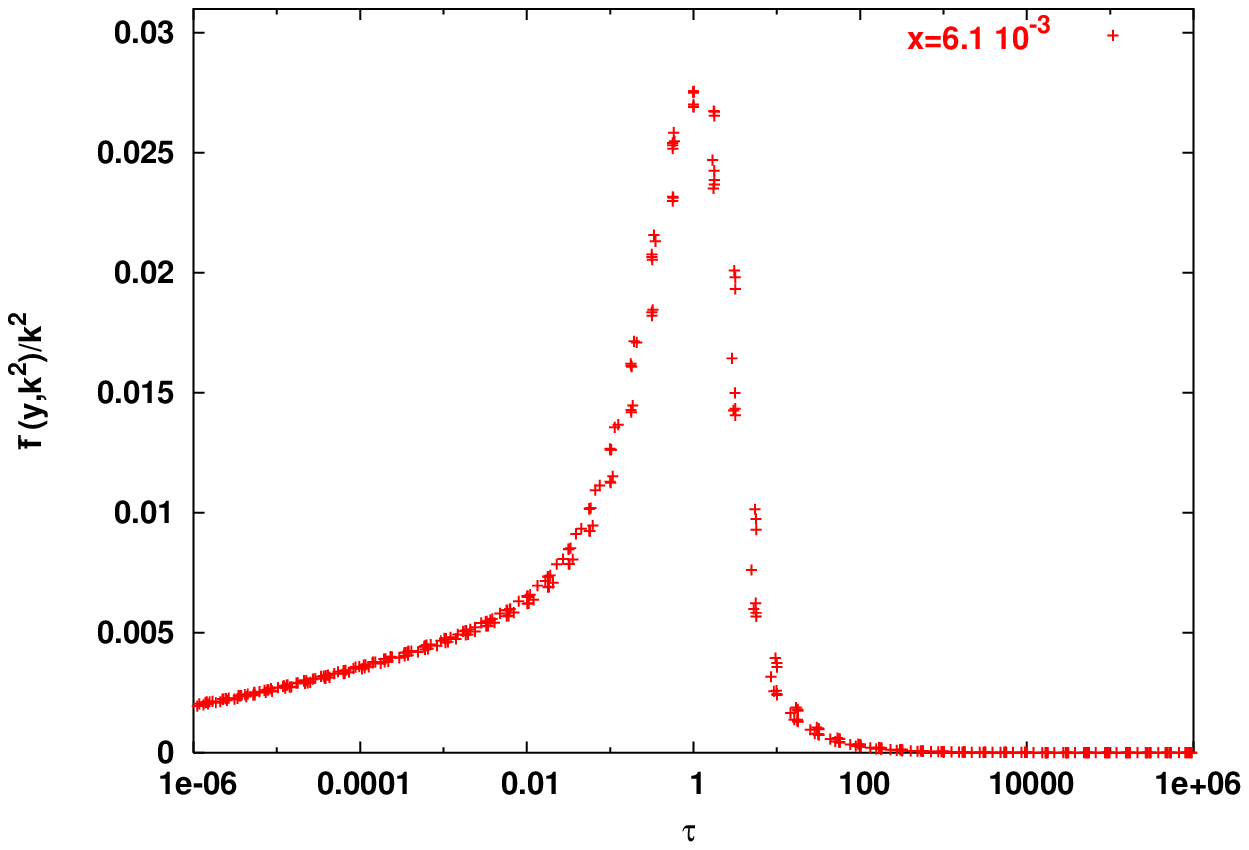,width=90mm} &
\psfig{file=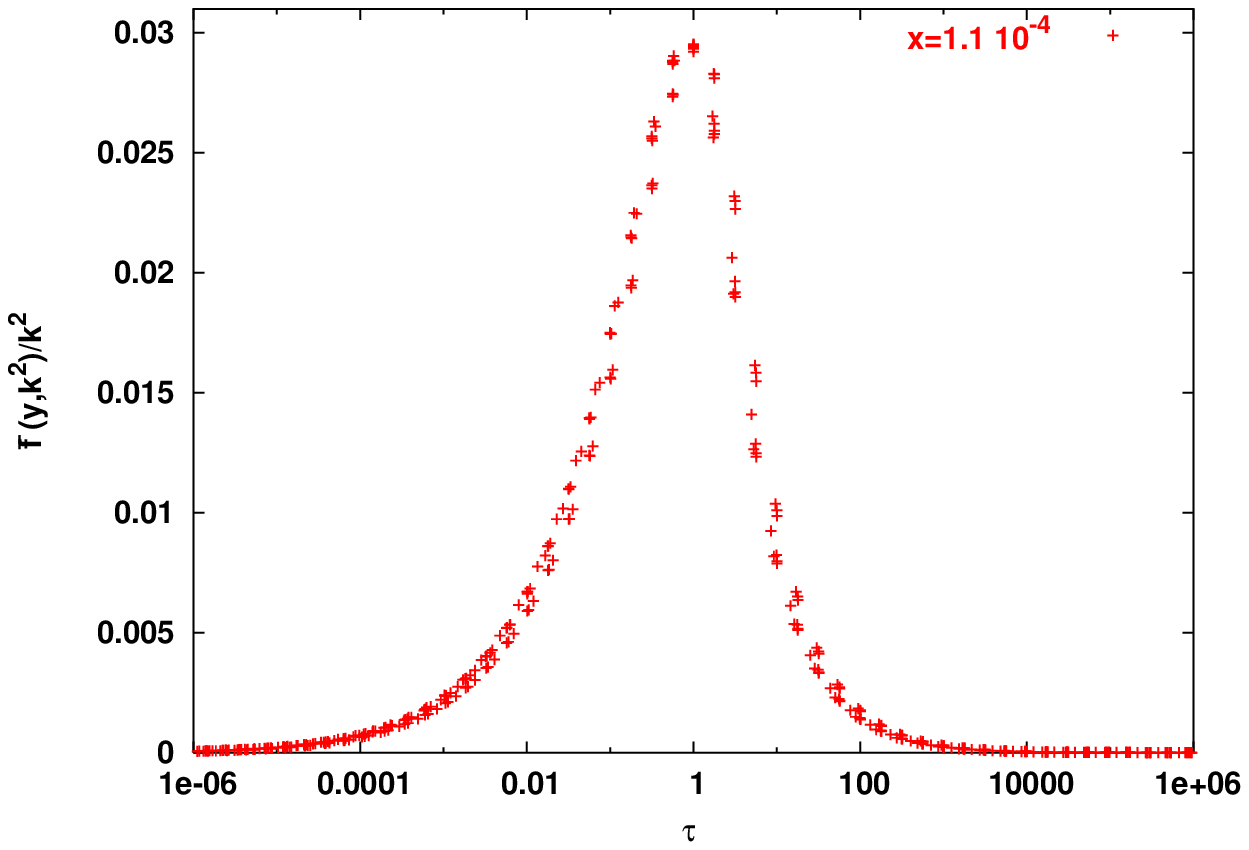,width=90mm}\\
 &  \\
\psfig{file=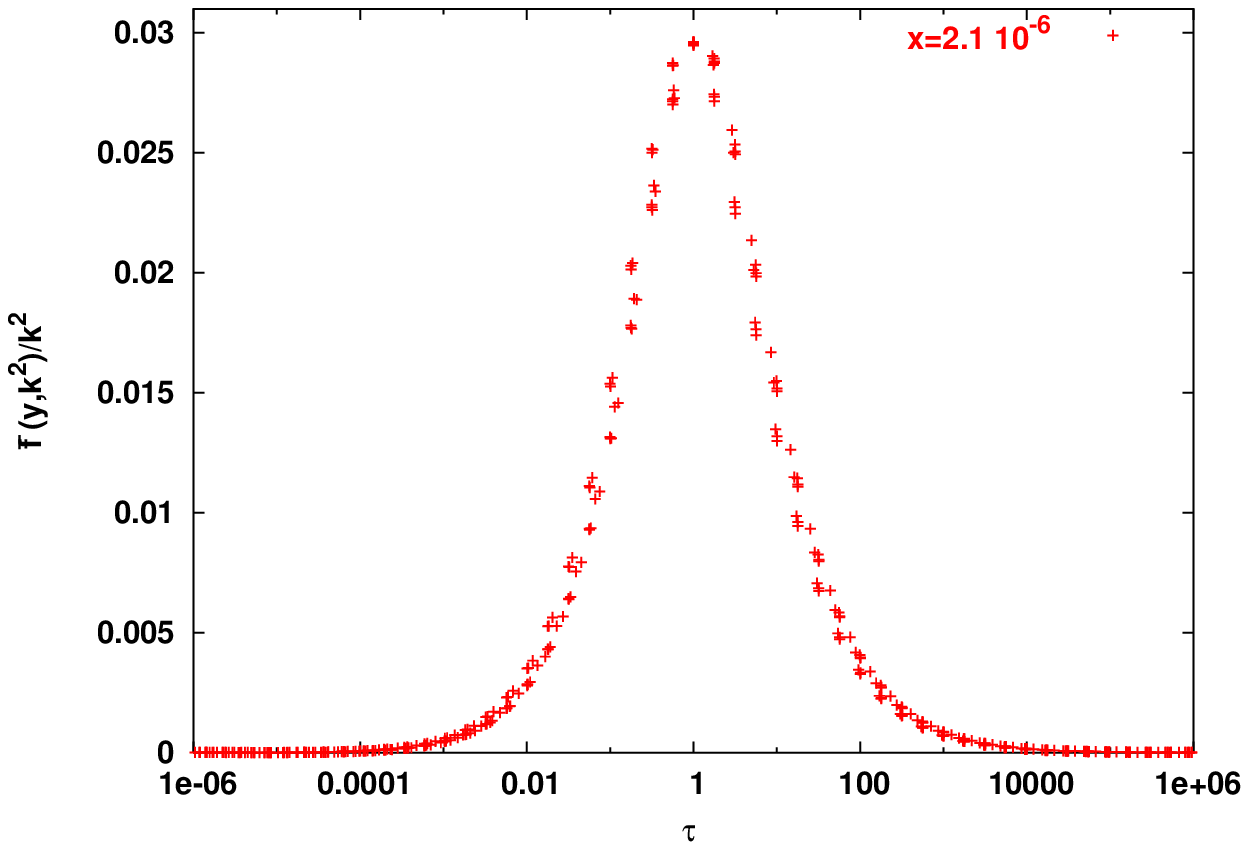,width=90mm} & \,
\psfig{file=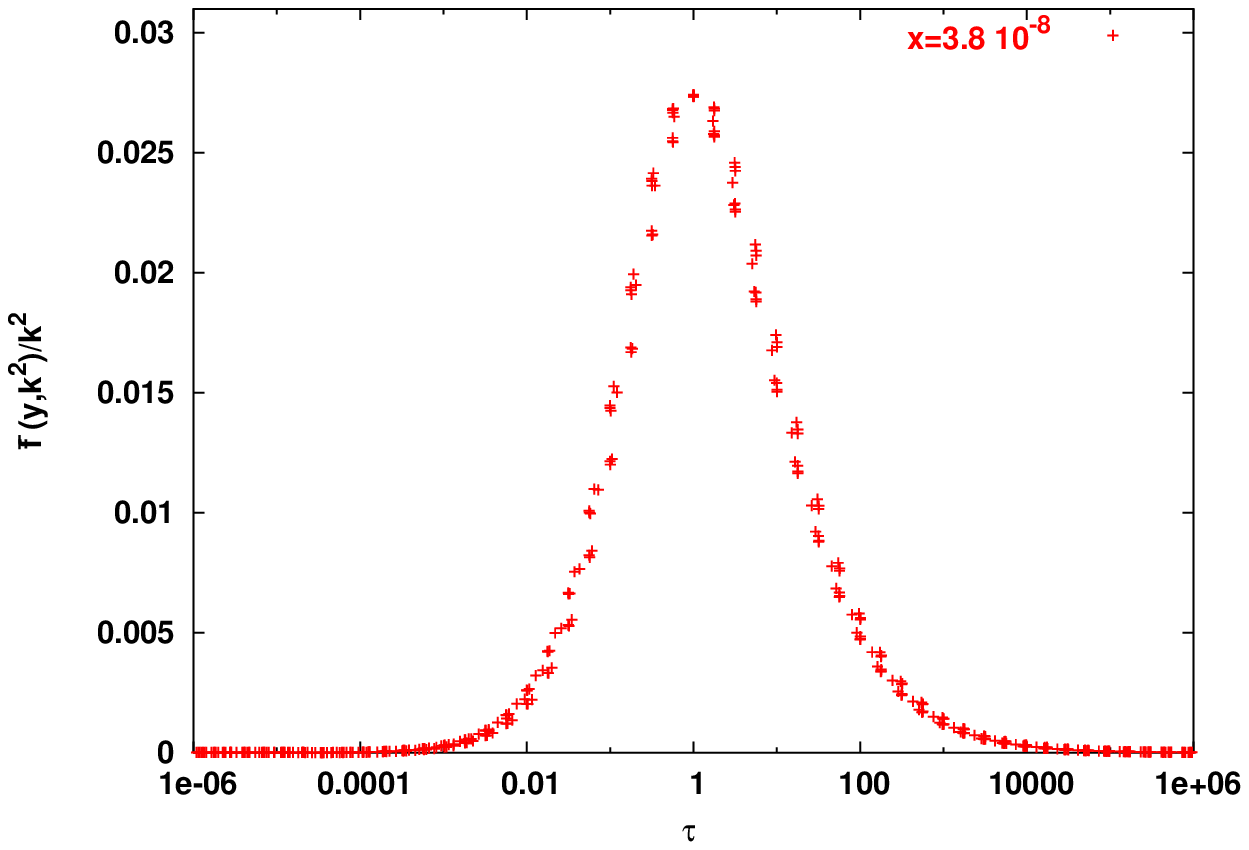,width=90mm}\\
 &  \\ 
\end{tabular}
\caption{\it Unintegrated gluon density function 
$\tilde{f}(x,k^2,b)\,/\,k^2$ as a function of scaling variable $\tau\,$ 
at fixed values of $x$ and different $b\,=\,0\,-\,12\,\,GeV^{-1}$
for the case of DIS on the proton. }
\label{Fig10}
\end{figure}

  With the help of $Q_{S}^{2}(b,x)$ it is easy to investigate scaling properties
of the unintegrated gluon density function $\tilde{f}(\ytau,k^2,b)$. Plotting
$\tilde{f}(x,k^2,b)\,/\,k^2$ 
as a function of only  $\tau\,=\,\frac{k^2}{Q_{S}^{2}(b,x)}$ at $b=0$, 
we obtain results presented in the Fig.\ref{Fig7-a}.
As it seems from the  Fig.\ref{Fig7-a}, the only approximate 
scaling of $\tilde{f}(x,k^2,b)\,/\,k^2$ exists,
if we collect all the data  at different values of $x$ .
This approximate scaling behavior is mostly pronounced near the maximum of the
$\tilde{f}(x,k^2,b)\,/\,k^2$ function and 
it is clearly broken in the "tales" of the function at large and small values
of $\tau$. Nevertheless, if we consider the $\tilde{f}(x,k^2,b)\,/\,k^2$
at fixed $x$ and at different $b$, then we see a perfect scaling behavior of 
the function, see Fig.\ref{Fig10}.
The only approximation scaling behavior of the unintegrated 
gluon density means, that $\tilde{f}(x,k^2,b)\,/\,k^2\,$ 
is likely a function of two variable, i.e. that
$\tilde{f}(x,k^2,b)\,/\,k^2\,=\,F(x,\tau(x,b))$
in the given framework.

\subsection{Saturation momenta for DIS process on the nuclei}

\begin{figure}[hptb]
\begin{tabular}{ c c}
\psfig{file=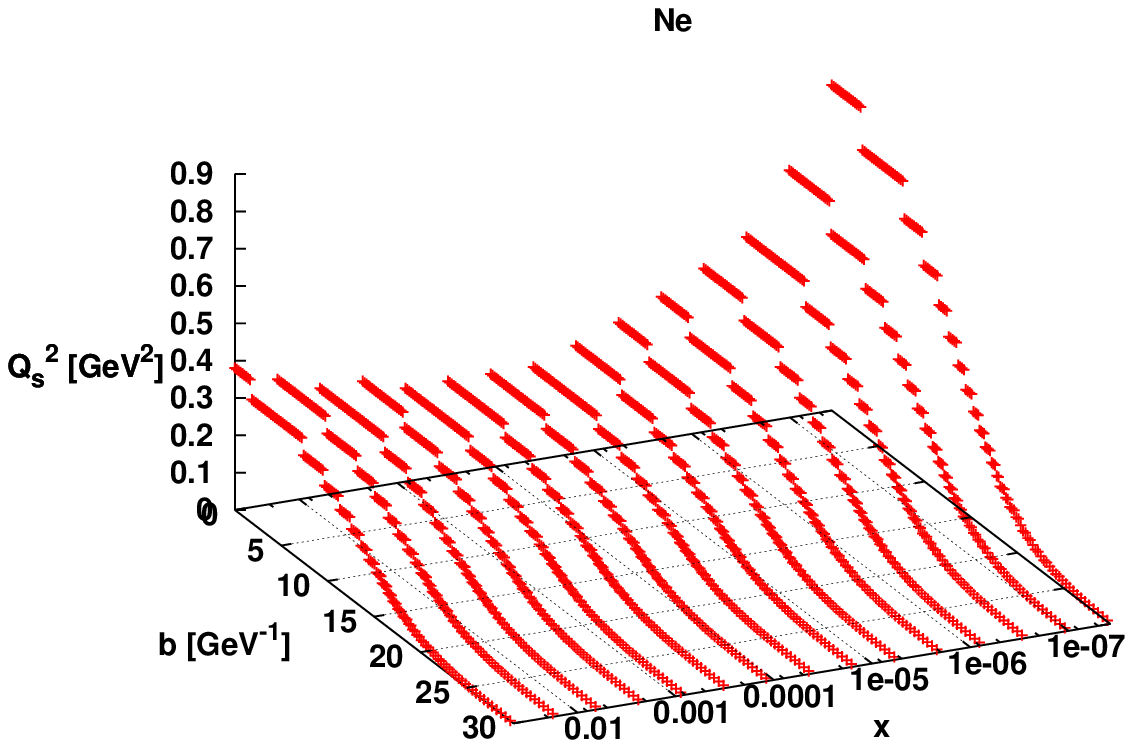,width=90mm} &
\psfig{file=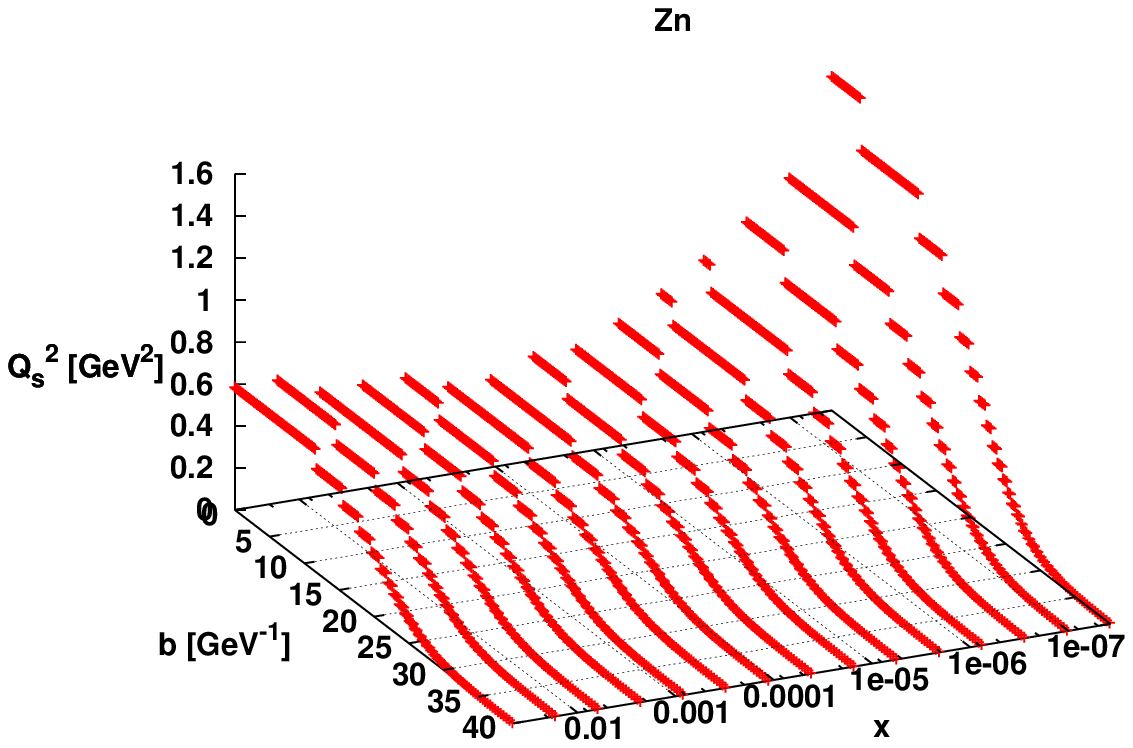,width=90mm}\\
 &  \\
\psfig{file=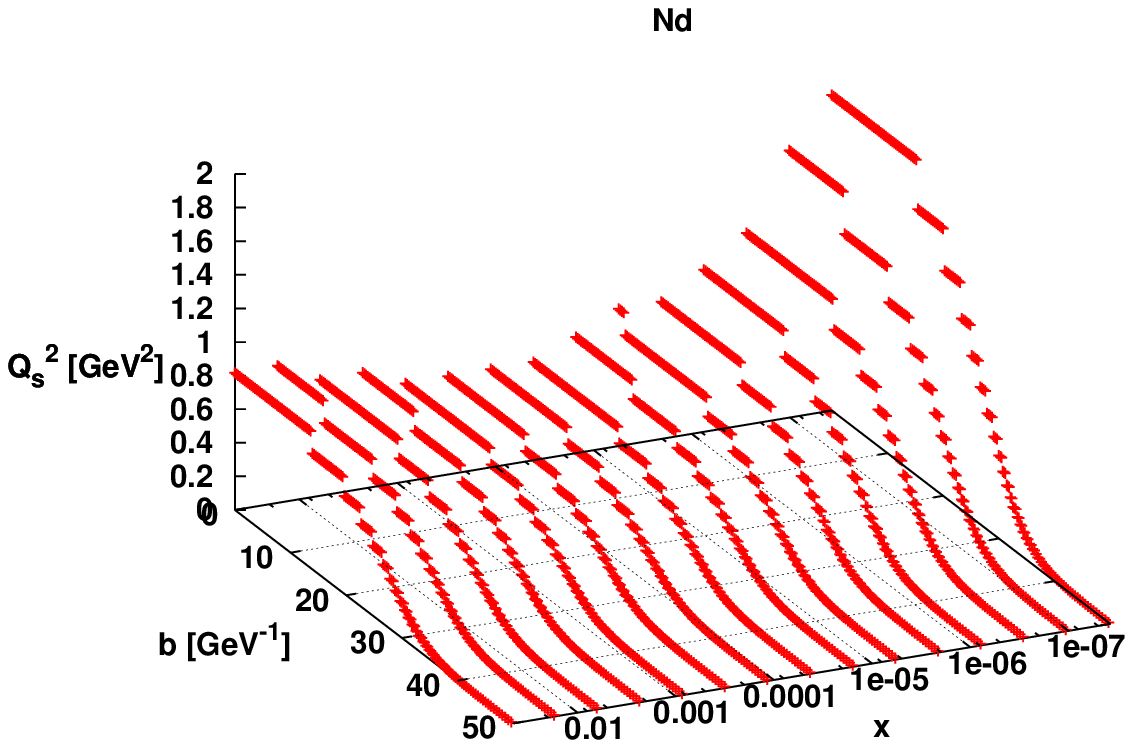,width=90mm} & \,
\psfig{file=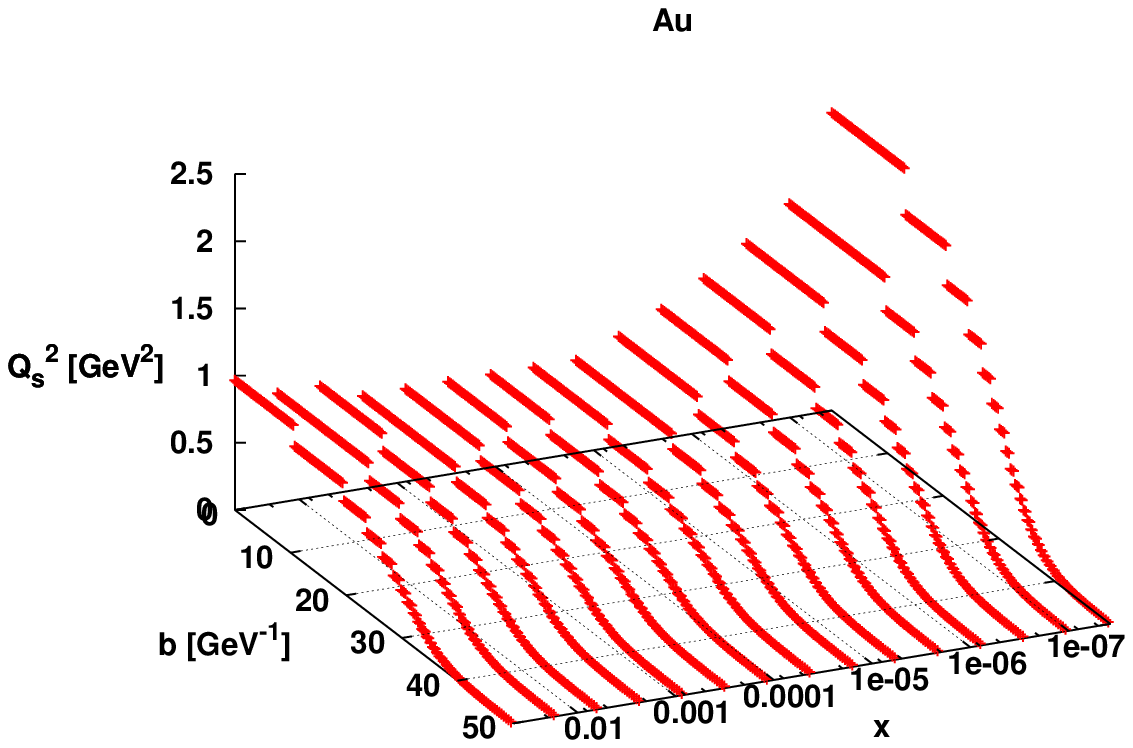,width=90mm}\\
 &  \\ 
\end{tabular}
\caption{\it Saturation momenta of nuclei as a function of x and
impact parameter. }
\label{Fig12}
\end{figure}

\begin{figure}[hptb]
\begin{tabular}{ c c}
\psfig{file=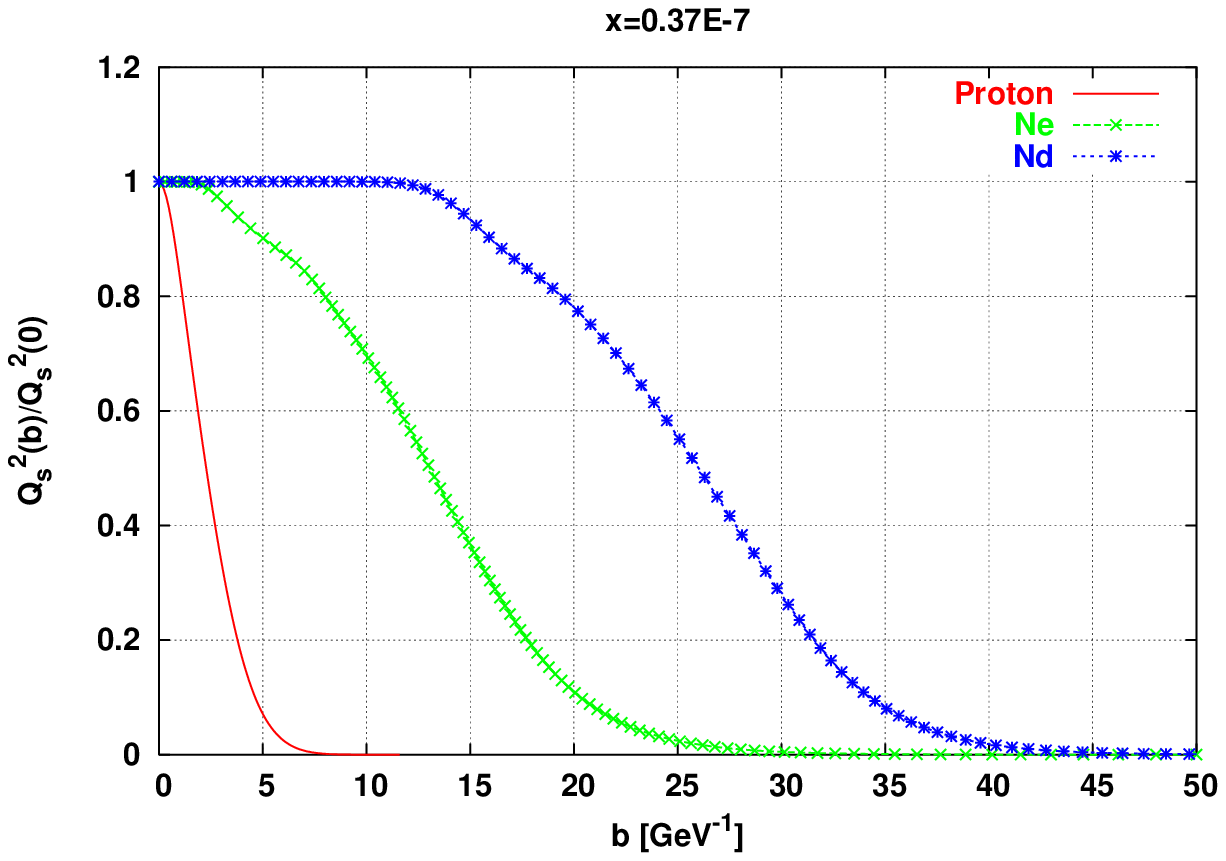,width=86mm} &
\psfig{file=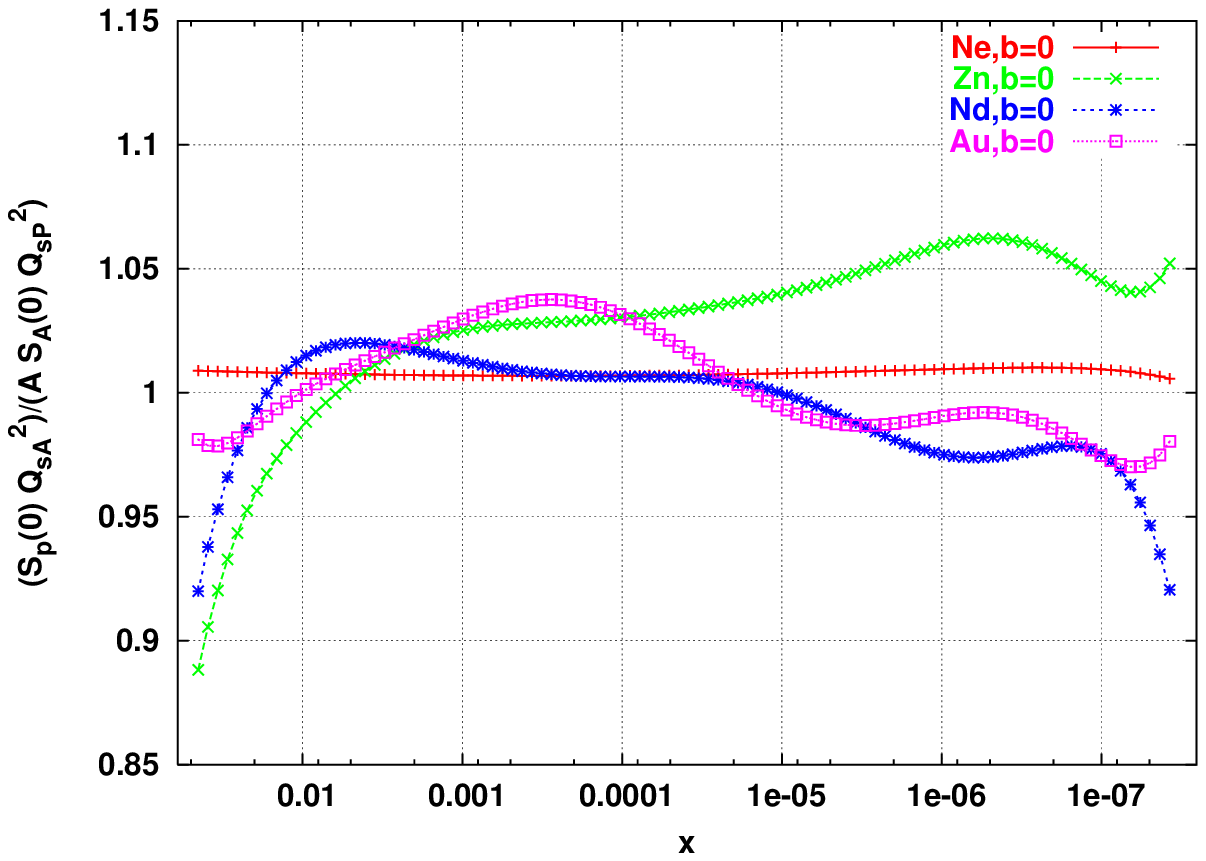,width=83.5mm}\\
 &  \\
\end{tabular}
\caption{\it Normalized impact parameter profile of the saturation
momenta of the proton and different nuclei in the left plot
of the picture and ratio Eq.\ref{SmN4} for the different nuclei in the right 
plot of the picture.}
\label{Fig12-b}
\end{figure}

The behavior of the saturation momenta of the different nuclei as
a function of impact parameter and x is presented in the 
Fig.\ref{Fig12} and in the left plot
of the Fig.\ref{Fig12-b}.
We parametrize the saturation momenta of the nuclei in the following form  
\beq\label{SmN1}
Q_{SA}^{2}(b,x)\,=\,Q_{S0A}^{2}(b)\,+\,Q_{S1A}^{2}(b)\,\Le\,
\frac{x_0}{x}\,\Ra^{d(b)}\,.
\eeq
The fitting of the data gives the following expression for the 
saturation momenta of the nuclei
\beq\label{SmN2}
Q_{SA}^{2}(b,x)\,=\,F_{SA}(x)\,A\,S(b)\,=
\Le\,8.84\,+\,12.56\,\Le\,
\frac{10^{-7}}{x}\,\Ra^{0.46}\,\Ra\,A\,S(b)\,\,\,GeV^{2}\,,
\eeq
where $S(b)$ is a Wood-Saxon nuclei profile function from Eq.\ref{F16}.
We see, that the coefficients in the $\,F_{S}(x)\,$
function in Eq.\ref{SmN2}
are the same as in the expression for the saturation momenta of the proton
in Eq.\ref{SmP2}. Indeed, if we
compare the expression  Eq.\ref{SmN2} with the  Eq.\ref{F15} we see,
that the coefficient $C$ from Eq.\ref{F15} must be the same as in Eq.\ref{F11}.
Therefore, apart the impact parameter dependence and number of nucleons
in target, the expressions for saturation momenta in Eq.\ref{SmP2} and 
Eq.\ref{SmN2} are the same.

 Now we could compare the expression \eq{SmP2} and \eq{SmN2}. It is easy to see
the following relation between saturation scales
\beq\label{SmN3}
Q_{SA}^{2}(0,x)\,\approx\,\frac{Q_{S}^{2}(0,x)\,A\,S_{A}(0)}{S_{p}(0)}\,,
\eeq
where $S_{A}(0)$ and $S_{p}(0)$ are the proton and nuclei profile functions correspondingly. So, 
from our expressions for the saturation momenta 
we obtain 
\beq\label{SmN4}
\frac{Q_{SA}^{2}(0,x)\,S_{p}(0)}{Q_{S}^{2}(0,x)\,A\,S_{A}(0)}\,\approx\,1\,.
\eeq
The independent calculations of this ratio presented in the right 
plot of the Fig.\ref{Fig12-b} shows that the expression \eq{SmN4}
is indeed correct. Now it is easy to obtain a $A^{n}$ behavior 
of the nuclei saturation momenta. Having in mind that
\beq\label{SmN5}
\frac{A\,S_{A}(b)}{S_{p}(b)}\,=\,A^{1/3}\,k_{A}(b)
\eeq
we obtain
\beq\label{SmN6}
Q_{SA}^{2}(b,x)\,=\,A^{1/3}\,k_{A}(b)\,Q_{S}^{2}(b,x)\,.
\eeq
We see, that we obtained a usual DGLAP $A^{1/3}$ dependence
of the nuclei saturation momenta with the additional modification coefficient
$k_{A}(b)$ which arises due the different sizes of proton and nuclei and which depend
on the atomic number $A$ of the nuclei. In the case of the considered nuclei this coefficient varies as 
$k_{A}(0)=0.28-0.34$. The DGLAP $A^{1/3}$ dependence
in \eq{SmN6} again underline the relative smallness of the shadowing effects in the processes of interest.
We also note, that  $k_{A}(b)$ coefficient
is related with the diluteness parameter $\kappa_{A}$ from \cite{Lev5} which 
is defined as
\beq\label{SmN7}  
\kappa_{A}(b)=\,A^{1/3}\,k_{A}(b)\,.
\eeq
For the considered nuclei this parameter varies as 
$\kappa_{A}(0)=0.76-2$ that support the point of view of \cite{Lev5}
on the nucleus as on the pretty dilute system of nucleons.

 Investigating the scaling properties of the 
nuclei unintegrated gluon density function
in the same way as it was done before for the proton case and 
again introducing the 
$\tau\,=\,\frac{k^2}{Q_{S}^{2}(b,x)}$ variable
we obtain results shown in the Fig.15.
\begin{figure}[hptb]
\begin{center}
\psfig{file=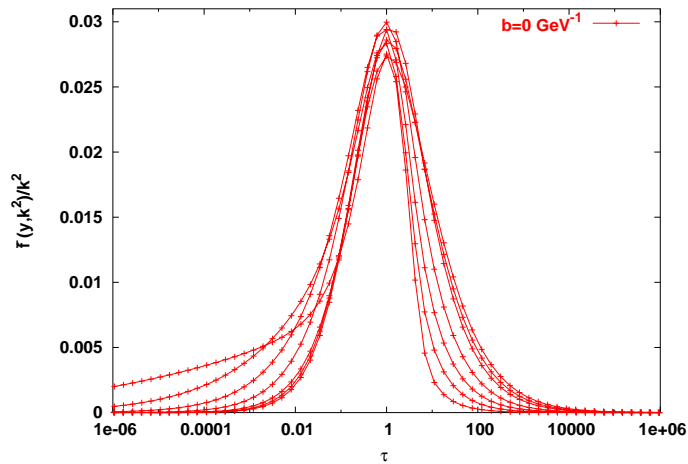,width=90mm} 
\caption{\it Unintegrated gluon density function 
$\tilde{f}(x,k^2,b)\,/\,k^2$ as a function of scaling variable $\tau\,$ 
at  $b=0$ and different $x\,=6.1\,10^{-3}\,-
\,3.8\,10^{-8}\,$ (curves from up to down in the left half of the plot) 
for the case of DIS on the gold ($A=197$).}
\end{center}
\label{Fig133}
\end{figure}
As in the case of the DIS on the proton we see, that
the scaling is precise only at fixed value of x. In general,
the scaling behavior of the unintegrated gluon density function is
only approximation, see again Fig.15 and  Fig.16.
\begin{figure}[hptb]
\begin{center}
\begin{tabular}{ c c}
\psfig{file=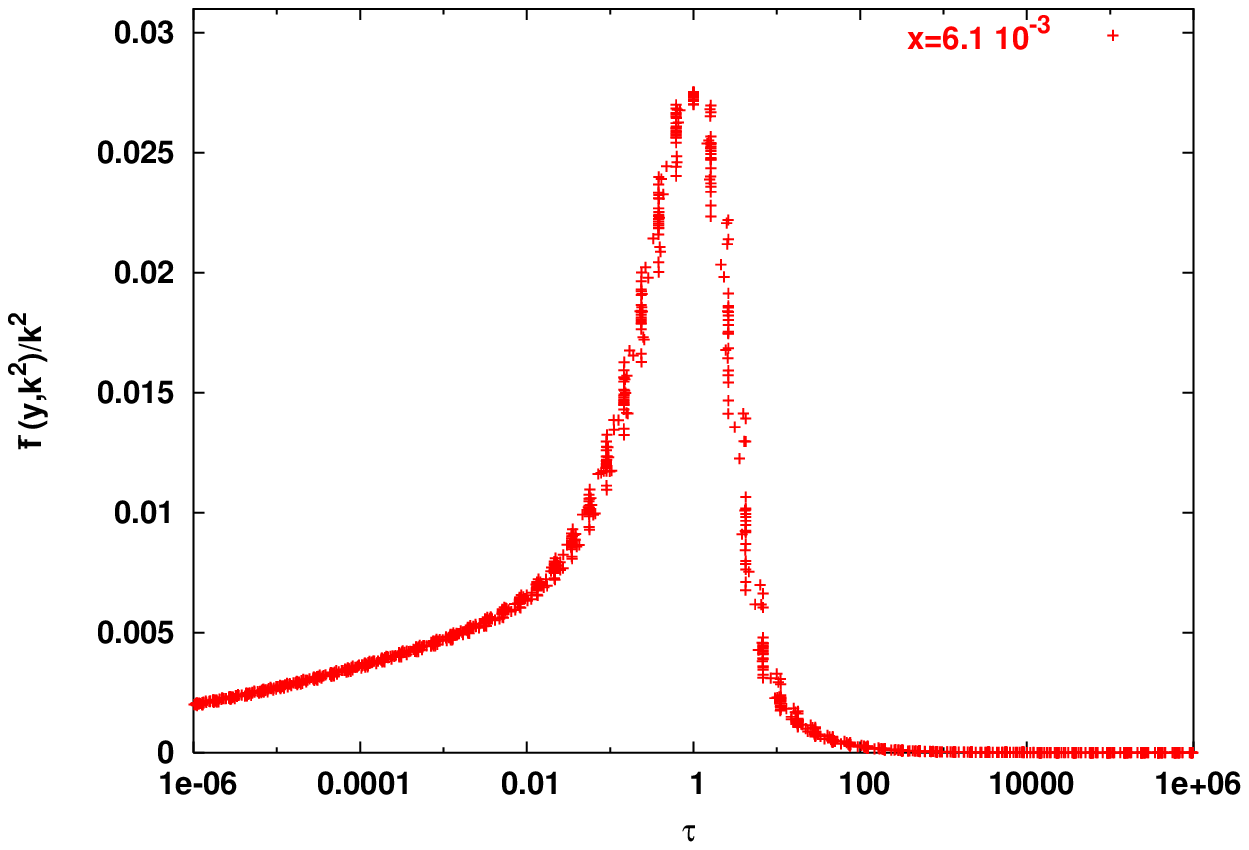,width=90mm} &
\psfig{file=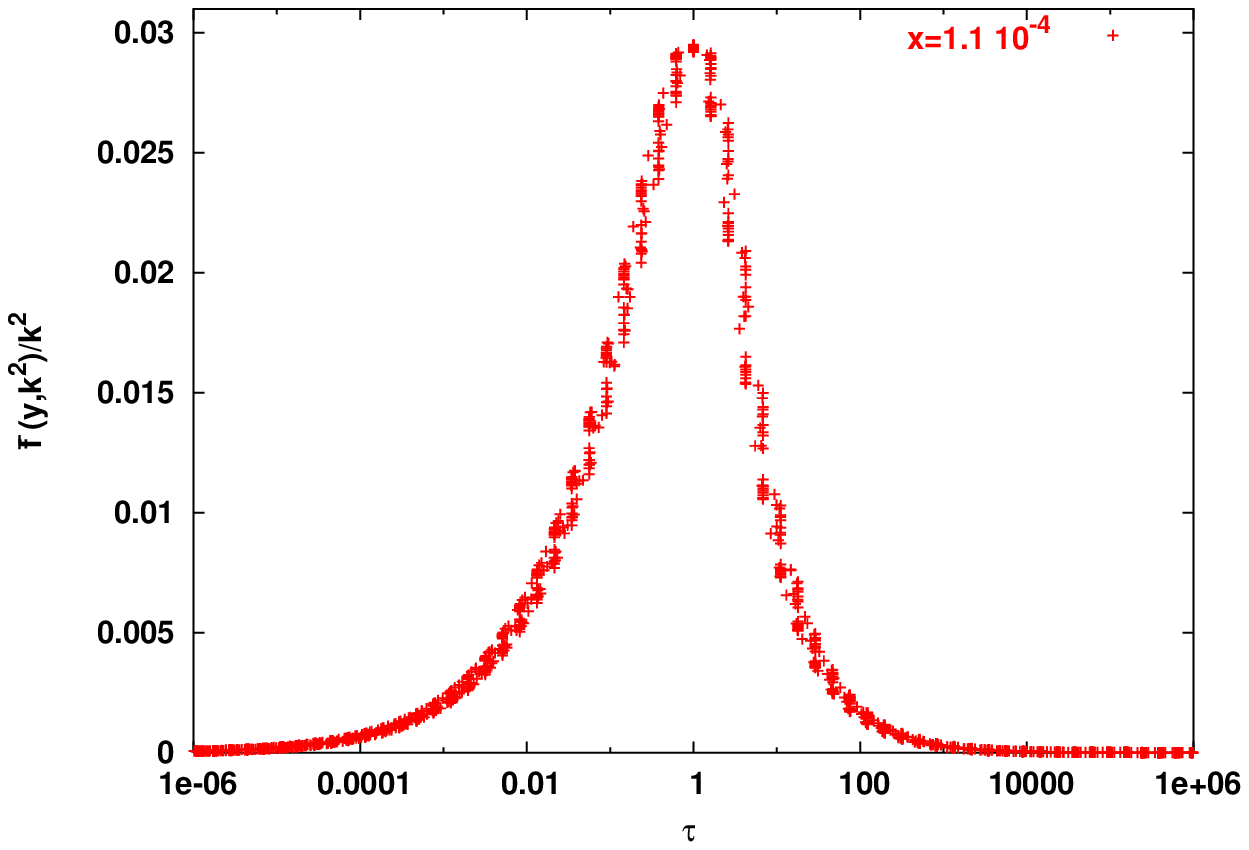,width=90mm}\\
 &  \\
\psfig{file=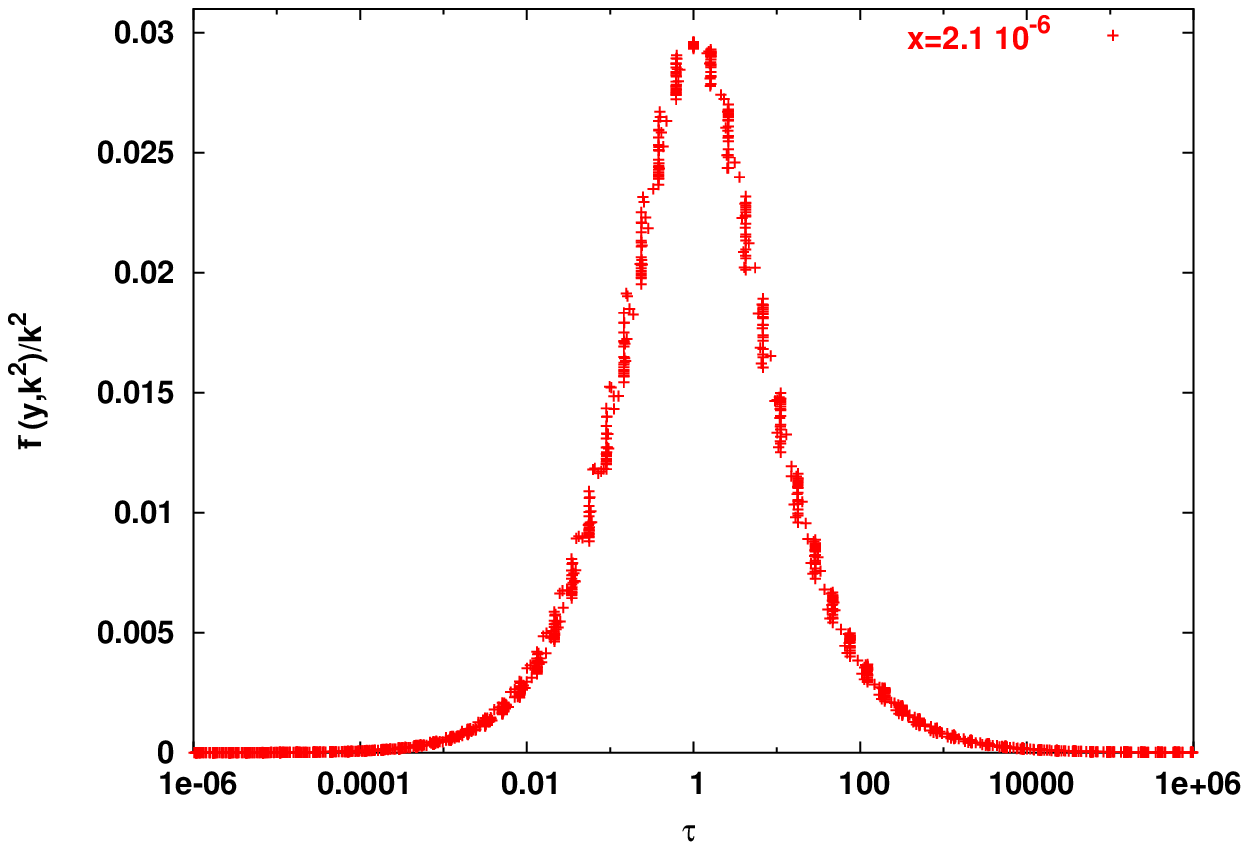,width=90mm} & \,
\psfig{file=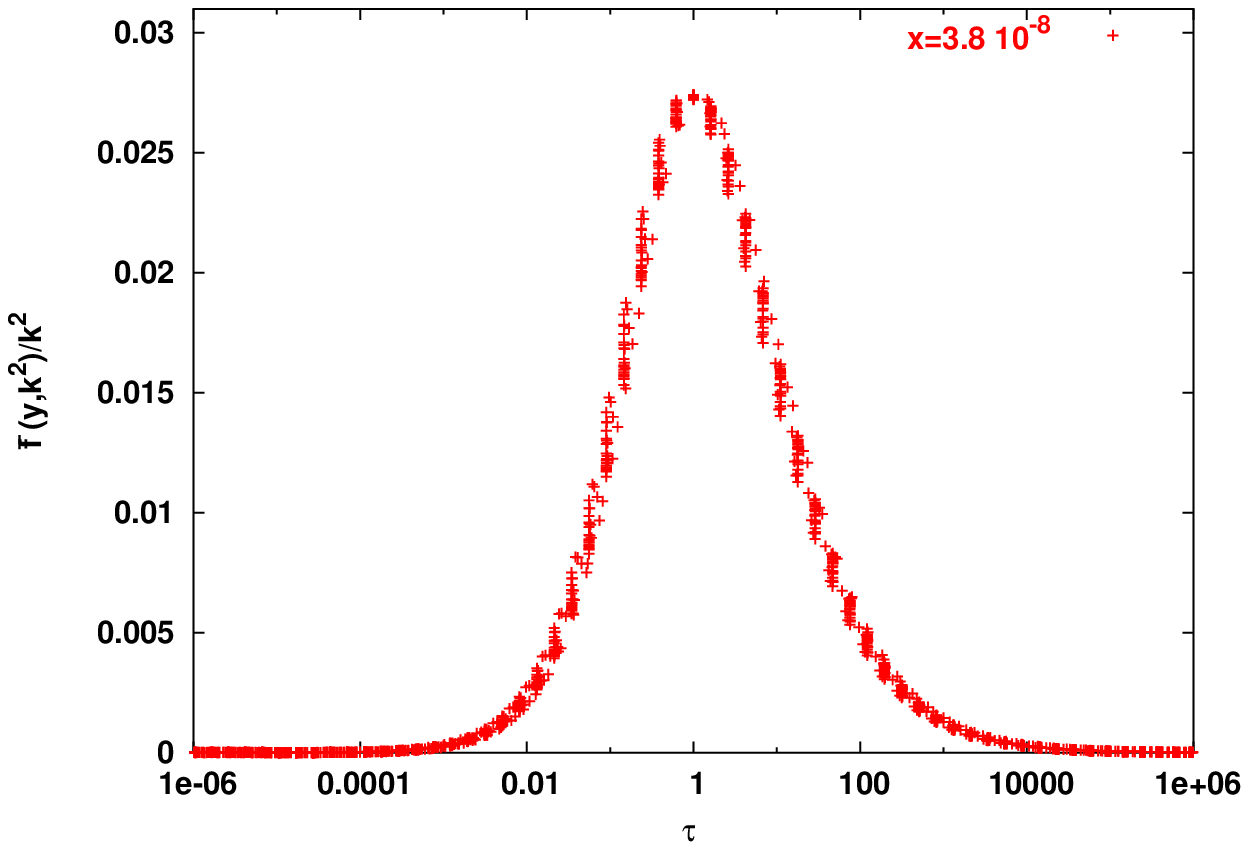,width=90mm}\\
 &  \\ 
\end{tabular}
\caption{\it Unintegrated gluon density function 
$\tilde{f}(x,k^2,b)\,/\,k^2$ as a function of scaling variable $\tau\,$ 
at fixed values of $x$ and different $b\,=\,0\,-\,110\,\,GeV^{-1}$
for the case of DIS on the gold ($A=197$). }
\end{center}
\label{Fig14}
\end{figure}

\section{Conclusion}

  As the main result of this paper we consider
the application of the BK evolution equation
with the local impact parameter dependence to the 
DIS processes on the nuclei at small values of Bjorken x
with the initial conditions of the rapidity evolution
similar to the usual GBW ansatz.
The precise form of the  initial conditions for
the case of DIS on the proton was
obtained in the paper \cite{Serg1} with the help of $F_2$
HERA data fit and
we considered a DIS process on the nuclei
using the same functional form of the initial conditions. 
Definitely,  it is not fully clear
why the main parameters of the initial conditions will not be changed if
the processes with nuclei instead the proton are considered. 
Nevertheless, surprisingly,
in the given framework we obtained the integrated gluon density function
which similar to the integrated gluon density functions from the  known parameterizations in 
the initial point of evolution,
see Fig.\ref{Fig1}-Fig.\ref{Fig111}. The explanation of this fact is very simple,
in the given framework the conditions for the applicability
of high energy formalism seems to be universal. 
Indeed, let's consider the $\alpha_{s}$ coupling
as a parameter which defines a physical "condition" of the process. 
We assume, that the value of $\alpha_{s}$ is determined by some averaged
saturation momenta. In considered range of x,
$x\,=4.5\,10^{-2}\,-\,3.8\,10^{-8}\,$, the value of saturated momenta and,
correspondingly the value of $\alpha_{s}$ is almost the same for the proton and nuclei in 
LO calculation scheme. It gives a 
physical explanation of this universality, the application 
of the BFKL and BK equations in DIS process is independent on the 
considered target of the process.

 It was mentioned above, that the interesting property of the obtained results 
for the integrated gluon density function
is that it stays in the range of
the results for the integrated gluon density functions obtained in the different parameterization 
schemes which base on the 
application of DGLAP equation to the low energy experimental data, 
see again Fig.\ref{Fig1}-Fig.\ref{Fig111}. This fact, justifying chosen form
of the initial conditions, causes a interesting question.
In general, there is no physical reason why the curves obtained in framework of BK equation
will be the same as the curves obtained in the DGLAP calculation scheme.
Clearly, first of all, the reason for the such behavior of the
curves  is that the LO BFKL
equation defines the same behavior for the integrated gluon density 
function as the LO DGLAP equation in some kinematic ranges. 
At large values of $Q^2$, when BK triple Pomeron vertex corrections 
are small, the coincidence of the curves is a sequence of this fact.
More interesting, that the curves 
obtained in BK equation formalism are still pretty close to the other curves
at smaller values of $Q^2$. 
As a explanation of this we assume that the triple
Pomeron vertex corrections are relatively small in the considered kinematic range
of x and $Q^2$.
As it seems, at given $Q^2$ and x the contribution of these corrections is not large, 
that makes resulting curves based on the BK and DGLAP equations
formalism to be similar. Going into the region of smaller  $Q^2$
at the small given and fixed values of x, we will see
a larger deviations between the results. At very 
small values of $Q^2$ the considered LO BK equation fails to describe data, 
that could be a sign not only of a significance of the triple Pomeron vertex 
corrections but also a sign of the need to add a further, "net" diagrams 
corrections into the process, see \cite{Serg1,bom4,bom,Bond,Bartels2,Bond1}. 
Of course, going into the region of smaller
x at fixed values of $Q^2$, we also will obtain the more significant deviation 
between the curves of different approaches caused by the large triple Pomeron vertex 
corrections.

 The additional mark of the smallness of the triple Pomeron vertex corrections
is a value of coefficient $\alpha_{A}$ from the Eq.\ref{Gd7}, see 
Fig.\ref{Fig2}. The LO DGLAP result gives $\alpha_{A}\,=\,1$ for any value of
$Q^2$. As it seems from Fig.\ref{Fig2}, even at $Q^2=2.5\,\,GeV^{2}$
and $x\propto\,10^{-3}$ the value of  $\alpha_{A}\,\approx\,0.8\,$
for the gold target. The obtained value is 
close to one and it is larger then the value of $\alpha_{A}$ obtained in \cite{Lev11}, 
for example. The same parameterization as in Eq.\ref{Gd7} we can write for the
$F_{2N}$ nuclei structure function, see  Eq.\ref{FS3}  and  Eq.\ref{FS4}.  
The calculations gives, that the $F_{2N}$ structure function is less sensitive
to the shadowing, i.e. the coefficient $\beta_{A}$ is closer to unity
then coefficient $\alpha_{A}$ in the same kinematical region.  

 The result of the anomalous dimension calculation for the case of the DIS 
on the proton,
see  Eq.\ref{An2} and Fig.\ref{Fig5}, shows a significant dependence of the
$\gamma_{P}$ on the value of $Q^2$. At large $Q^2$ and at small values of x,
the  $\gamma_{P}\,\rightarrow\,1/2\,$, as it must be in the case of BFKL
asymptotic results. For the smaller values of $Q^2$ the anomalous dimension is 
larger then half, that shows a corrections arise due the triple Pomeron vertex.
The same calculations for the nuclei, see Fig.\ref{Fig6}, show a dependence of the
value anomalous dimension of the nuclei on the atomic number A. This dependence especially 
underlined for the case of the small values of $Q^2$, in contrary to the results 
of calculations in \cite{Lev11}.

 Another problem, considered in the paper, is a  
problem of the saturation momenta in the 
DIS on the proton and on the nuclei. First of all, initially
we assumed a similar and factorized form for the saturation momenta of the proton and 
of the nuclei in the process of interest, see Eq.\ref{F11} and Eq.\ref{F15}.
The same factorized form was preserved in the final
expressions for the saturation momenta of the proton and of the nuclei
after the evolution over rapidity, see Eq.\ref{SmP2} and Eq.\ref{SmN2},
that in some sense justify the calculations of  \cite{Lev11,Lev2,Lev3} 
for example.
Nevertheless, there is a principal difference 
between the calculations of the \cite{Lev11,Lev2,Lev3} 
and present one which must be underlined.
In spite of the introducing the form of impact parameter dependence in
the solution of evolution equation after evolution, as it was done in 
\cite{Lev11,Lev2,Lev3}, 
the actual form of impact parameter dependence
of the saturation momenta in the present framework 
is calculated through the rapidity evolution
and, therefore the form of the saturation momenta with impact parameter dependence
is different from the used in \cite{Lev11,Lev2,Lev3}.

 The performed  calculation of the saturation scale shows,
that we did not observe
a suppresion over atomic number in the expression for the saturation momenta,
which was obtained in the similar calculations in different papers, see for example
\cite{Lev11, Armesto1,Compar1,Compar3} and references therein, see \eq{SmN6}
and plot in Fig.\ref{Fig12-b}. 
Still, we used a definition
of saturation momenta different from the definition of \cite{Lev11}, 
but our results are also different from the results of \cite{Armesto1}
where the similar definition was used. The first simple fact,
which could explain this result, 
is that we obtained and used a different expression
for the saturation momenta. Our expression contains two parts,
one part of this expression is a
constant and second part is a function which grows with rapidity.
This future of the considered model may be explained by the  local
impact parameter dependence introduced in the calculations.
From the Fig.9-Fig.\ref{Fig12} we see, that in the large range 
of $x$ the saturation momenta
is almost a constant, the growth became at $x\propto\,10^{-5}$ only.
Considering the saturation momenta as the characteristic momenta of the scattering system
which related with the averaged parton density
of the system, we see, that until  $x\propto\,10^{-5}$ this density does not 
grow so fast. 
Therefore, the flatness of the saturation momenta as the function of $x$
in broad small $x$ region
might be explained by the linear growth of 
target area in the impact parameter space accounted by the introduced impact parameter dependence.
When a speed of increase of the area of the target 
is similar or larger than a speed of increase of the characteristic momenta then
the saturation scale does not change so much. Only when the grow
of the momenta is larger than the grow of the area of the target, only then
we observe a increase of saturation momenta with the increasing of energy.
We can conclude, therefore, that interplay between the linear growing
of the area of the target in the impact parameter space and growing
of the parton density "delays" growing of the saturation momenta.
Only at asymptotically large energies, when the effect of the growth of the target
area will be negligibly small, the expected suppression over the
atomic number $A$ perhaps could be observed .

 Relating the proton and nuclei saturation momenta, see expression \eq{SmN6},
we introduced a parameter $k_{A}(b)$ which is a parameter of the diluteness
of the system, see \cite{Lev5}. As it was underlined in \cite{Lev5}, see also
references therein, this parameter is not so large for the case of real nucleus.
Without discussing a new approaches introduced in \cite{Lev5}
and a need of the additional rescattering corrections to the
system amplitude, see for example \cite{Bond}, we can conclude
nevertheless, that
the relative smallness of this parameter shows on the smallness
of the triple vertex (shadowing) corrections as well.

 It is also interesting to compare obtained results
with the 
other results on the saturation momenta. First of all, it is instructive to compare
obtained form of the proton saturation momenta with the result
of the GBW paper \cite{GolecBiernat1}, see Fig.\ref{Fig7-b}. As it seems
from the plot, the compared saturation scales have different 
functional form and also, as a consequence, different numerical values
at different values of $x$. At relative large values of $x$ the obtained in this paper saturation 
scale is larger then the GBW one, whereas at very small values of 
$x$ the  GBW is larger. The explanation of this result
was done before, the difference arises mainly due the very different
functional form of the functions. The expression \eq{SmP2}  have  a constant 
part which does depend on x, which arises because of the impact parameter dependence,
and, therefore, the growth of the scale with the $x$ decreases begins much later then 
in the GBW scale case. Therefore,  it is difficult to compare our result with the 
different result based on the GBW scale parameterization and geometrical scaling 
application, see \cite{Compar1} for example.
There the GBW saturation scale parameterization was used in order to extract the 
$A^n$ dependence of the nuclei saturation momenta.  Obtained in these calculations
coefficient $n$ is different from DGLAP $n=1/3$ value obtained here, as it may be expected simply because 
of the different saturation scales parameterizations.
Nevertheless, in the contrast to the results of \cite{Compar1} in the 
paper \cite{Compar2} the similar value of the coefficient $n=1/3$ for the nuclei saturation 
scale was obtained after the light nuclei data excluding  from the fitting procedure.
Still, it must be underlined, that the compared models have a different
calculation frameworks, that makes they comparison pretty difficult, see Appendix A
for some remarks on this subject.

 Obtained through the calculation expressions of the saturation momenta  
allow to investigate the scaling properties of the unintegrated gluon density function.
In the given framework and in the considered kinematical
region we found that the scaling is only approximate future of the function
as it seems from the Fig.\ref{Fig7-a}-Fig.\ref{Fig10} and 
Fig.15-Fig.16. In general
the unintegrated gluon density function depends on the two variables,
scaling variable $\tau(b,x)$ and on the energy of the process 
(rapidity or value of x). 
Nevertheless, the scaling is precise when we fix the energy and consider
different values of impact parameter. In this case, instead
the different solutions at different impact parameters, the scaled solution
arises and its depends only on the $\tau(b)$ variable.

 Finally we would like to underline, that given calculations we consider as a 
first step
in establishing of the framework for the 
investigation of more complex processes at very small values of x such as 
amplitude of proton-proton scattering at LNC energies or calculations
of the amplitude of the exclusive Higgs boson production. 
The demonstration that in the framework of the interacting BFKL pomerons it is 
possible to describe a bulk of the
DIS data on the proton and nuclei,  leads us towards the
application of the model in the calculations of these complex processes at the  LHC collider energies.


\section*{Acknowledgments}

\,\,\,S.B. especially grateful to Y.Shabelsky for the  discussion
on the subject of the paper.
This work was done with the support of the Ministerio de Educacion
y Ciencia of Spain
under project FPA2005-01963 together with Xunta de Galicia
(Conselleria de Educacion).


\newpage
\section*{Appendix A:}

\renewcommand{\theequation}{A.\arabic{equation}}
\setcounter{equation}{0}

 In this appendix we would like to come back to some details concerning
the DIS process on the proton from \cite{Serg1}. The main problem, which 
we faced in that calculations is a fail of the description
of small $Q^2\,<\,1\,\,GeV^2$ DIS data for $F_2$ function, see
details in \cite{Serg1}.
If we will compare our approach with the saturation approach from 
\cite{GolecBiernat3,Kowalski1,Kowalski2}, which describe all data
of DIS on the proton, we immediately will see that we missed a correct 
description of gluon density function at small values of momenta, i.e.
we do not reproduce the initial condition of DGLAP evolution
equation for the DIS process. This result may be explained
by the absence of the additional "net" diagrams in the BK approach.
Indeed, when in the DIS process $Q^2\,<\,Q_{S}^2$ then the use of only
"fan" diagrams is acceptable. But when $Q^2\,>\,Q_{S}^2$,
i.e. when the size of projectile is large then the averaged
size of partons in the proton, then we definitely miss
additional contributions in the amplitude which represented by the
"net" diagrams, see \cite{Bond1}. 
Taking into account that 
we obtain for the proton saturation momenta
$Q_{S}^{2}\,\approx\,1\,GeV^2$ we see that indeed
we have to face a problems when $Q^2\,<\,1\,\,GeV^2$.

 Another hint to the solution of this problem is that during the 
calculations we obtain that the triple pomeron corrections
are relatively small, see also \eq{SmN7} and \cite{Lev5}.
Therefore, in spite of the long small $x$ evolution we,
as it seems,  missed a 
correct initial DGLAP type condition in the region of small
$Q_{S}^{2}$ which is crucially important in this kinematic region.
So, as a resolution of the problem, it will be interesting
to combine the DGLAP type initial conditions from
the  \cite{GolecBiernat3} with the BK evolution equation with local impact 
parameter dependence. This task we will leave for another publications.

\begin{figure}[hptb]
\begin{center}
\psfig{file=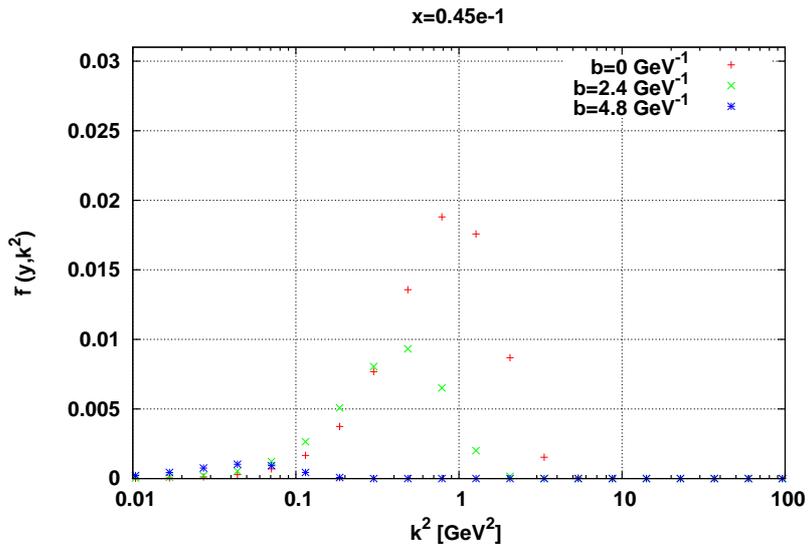,width=110mm} 
\caption{\it Initial profile of the unintegrated gluon density function 
$\tilde{f}(x,k^2,b)\,$ as a function of momenta $k^2\,$ 
at fixed values of $x$ and different values of $b$
for the case of DIS on the proton..}
\end{center}
\label{AppA1}
\end{figure}
\begin{figure}[hptb]
\begin{center}
\begin{tabular}{ c c}
\psfig{file=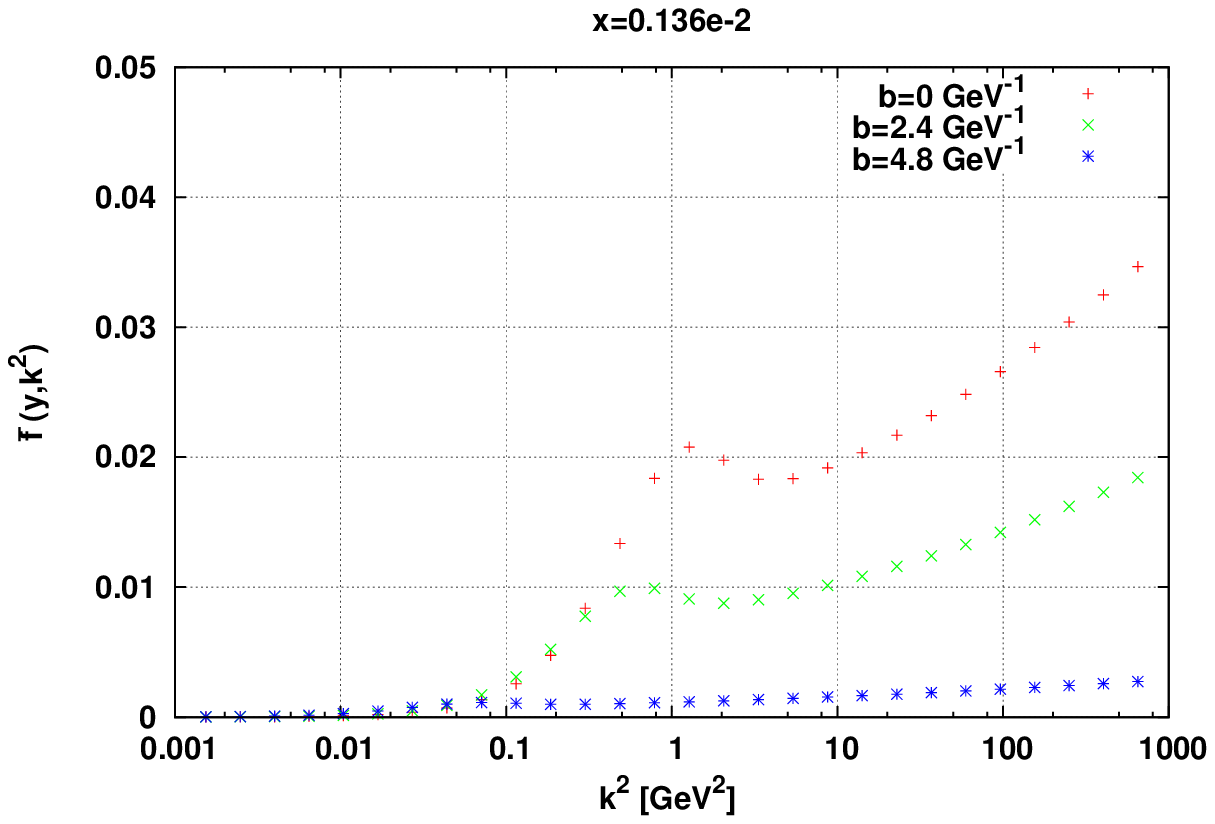,width=90mm} &
\psfig{file=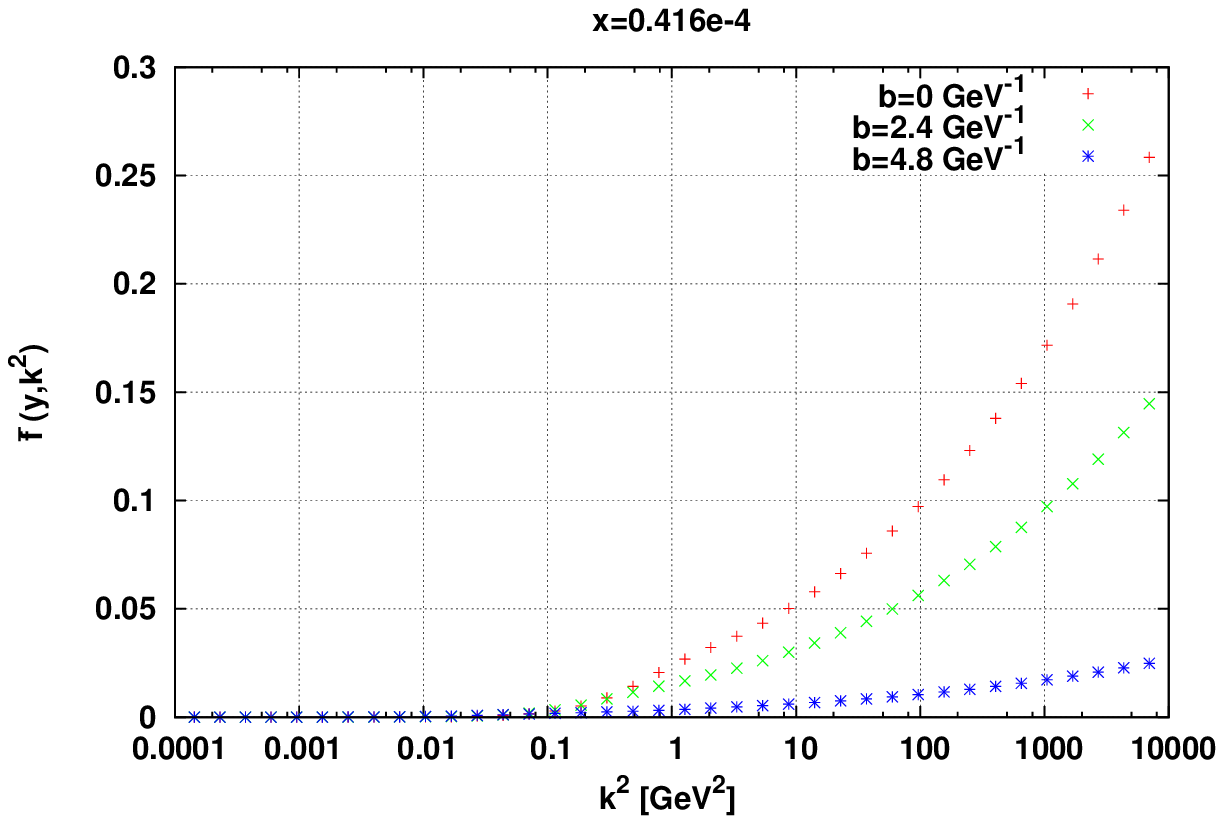,width=90mm}\\
 &  \\
\psfig{file=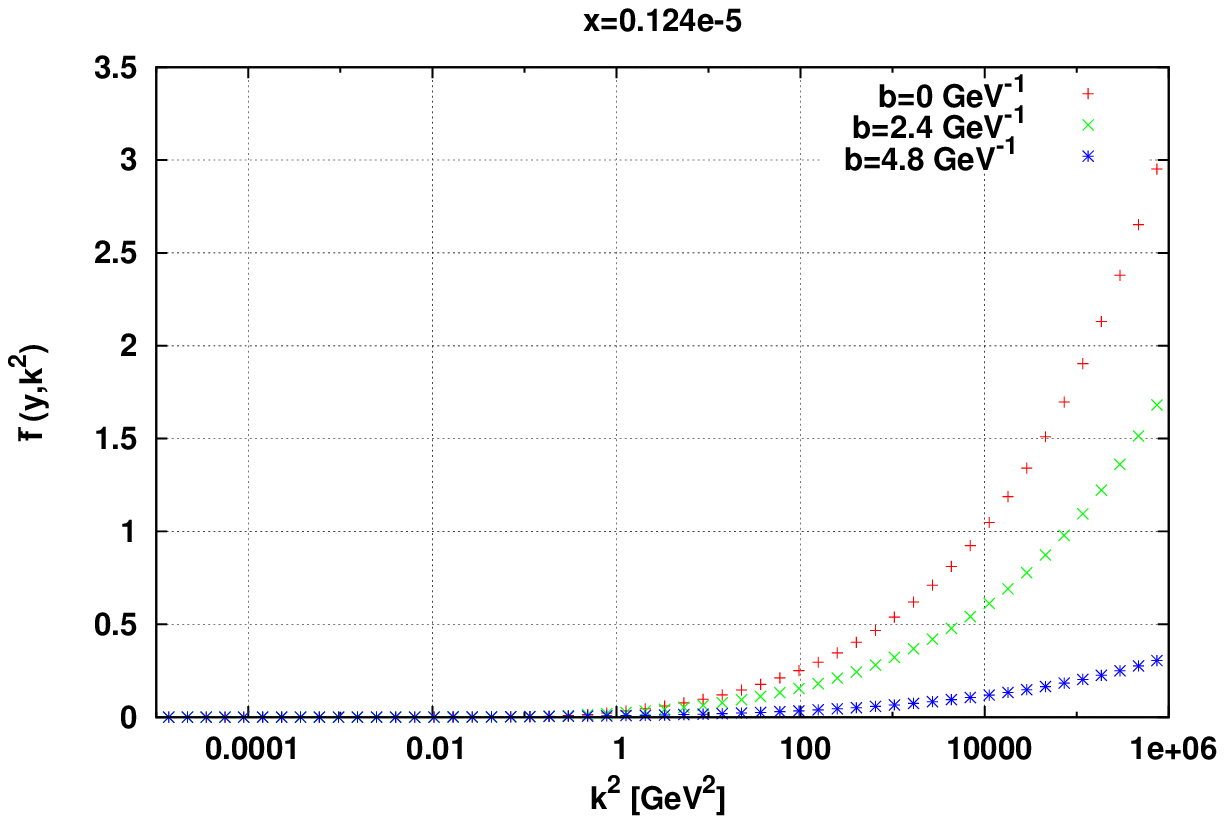,width=90mm} & \,
\psfig{file=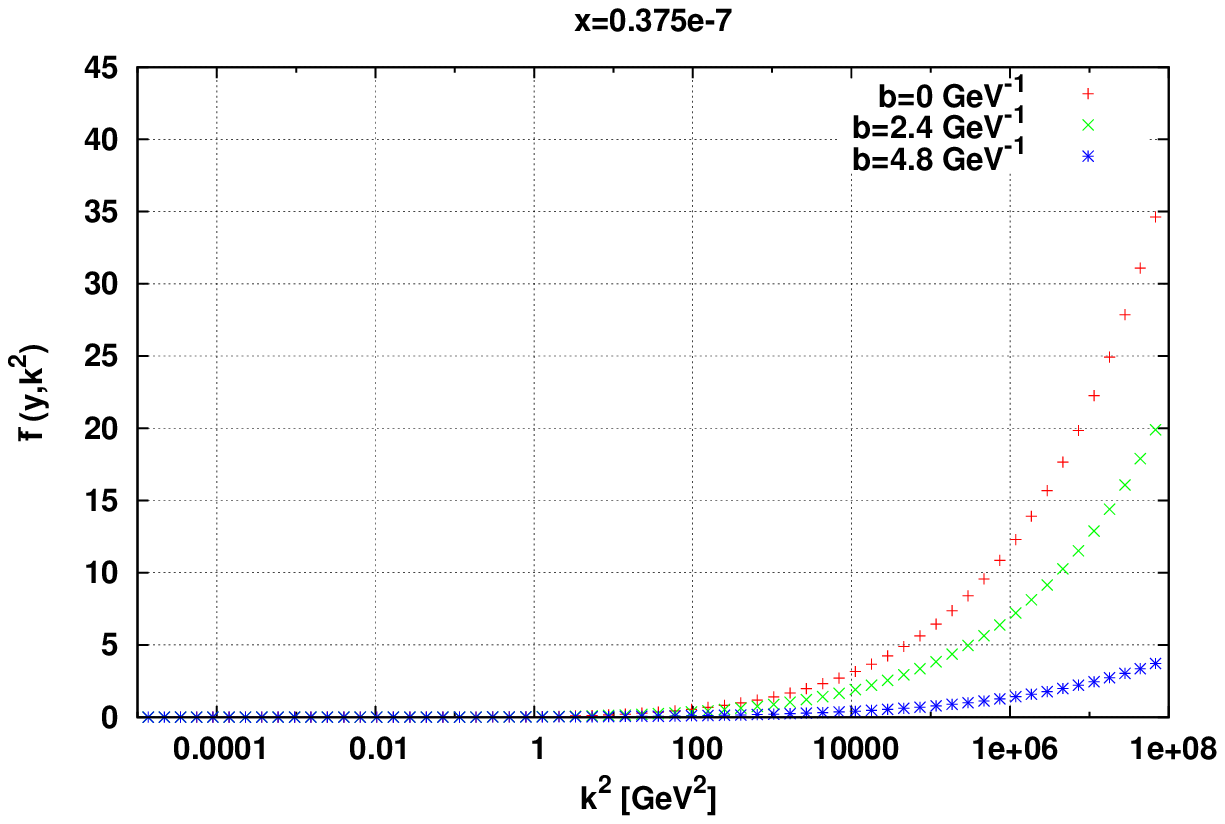,width=90mm}\\
 &  \\ 
\end{tabular}
\caption{\it Unintegrated gluon density function 
$\tilde{f}(x,k^2,b)\,$ as a function of momenta $k^2\,$ 
at fixed values of $x$ and different values of $b$
for the case of DIS on the proton.}
\end{center}
\label{AppA}
\end{figure}

 As a illustration of the difference between the present model and
model of \cite{GolecBiernat3}
we present here a plots
of the unintegrated gluon density function \eq{F4} as a function of momenta
at different values of $x$ and impact parameter.
This function has a dimension 
(as opposed to the dimensionless $\tilde{f}(x,k^2,b)\,/\,k^2$ function), 
therefore the numerical values of the
function in the plots is not important and we also skip the dimension of the function
on the plots.
Nevertheless, it is interesting to trace a creation of the large $k^2$
tale of this function beginning from the initial condition, Fig.\ref{AppA1},
and through the small $x$ evolution, Fig.\ref{AppA}.
Comparing this plots with the similar ones from the \cite{GolecBiernat3}
we see that our functions is absolutely different in the region of
large $k^2$, that may be explained as a result of the small $x$ evolution.
But there is a doubt that this tail explains a fail of the approach 
in the description of the $F_2$ function in the region 
of small $Q^2$. This large $k^2$ tail affects on the exclusive
processes mainly, whereas the inclusive ones are sensitive to the region of 
small $k^2$ which seems to be not so different from the presented in 
the \cite{GolecBiernat3}.


\end{document}